\documentclass[11pt]{article}
\usepackage{amsfonts}
\usepackage{amssymb,amsmath,mathrsfs}
\usepackage{graphicx,picins}
\usepackage{subfigure}
\usepackage{float}
\usepackage[numbers,sort&compress]{natbib}
\allowdisplaybreaks[4]

\begin{document}
\title{\bf{Integrable nonlocal complex mKdV equation: soliton solution and gauge equivalence} }
\author{ Li-Yuan Ma$^{a}$, Shou-Feng Shen$^a$ and Zuo-Nong Zhu$^b$\footnote{Corresponding author. Email: znzhu@sjtu.edu.cn; mathssf@zjut.edu.cn; mly2016@zjut.edu.cn}
\\
$^a$ Department of Applied Mathematics, Zhejiang University of Technology,\\ Hangzhou 310023, P. R. China\\
$^b$ School of Mathematical Sciences, Shanghai Jiao Tong University,\\ 800 Dongchuan Road, Shanghai, 200240, P. R. China}
\date{ }
\maketitle

\begin{abstract}
 In this paper, we prove that the nonlocal complex modified Korteweg-de Vries (mKdV) equation introduced by Ablowitz and Musslimani [Nonlinearity, 29, 915-946 (2016)] is gauge equivalent to a spin-like model. From the gauge equivalence, one can see that there exists significant difference between the nonlocal complex mKdV equation and the classical complex mKdV equation. Through constructing the Darboux transformation(DT) for nonlocal complex mKdV equation, a variety of exact solutions including dark soliton, W-type soliton, M-type soliton and periodic solutions are derived.
\end{abstract}

%PACS number(s): 05.45.Yv, 02.30.Ik\\
\section{Introduction}
\qquad The KdV equation $q_{t}+ 6 qq_x+q_{xxx}=0$ and the mKdV equation $q_{t}+ 6 q^2 q_x+q_{xxx}=0$ describe the evolution of small amplitude and weakly dispersive waves which occur in the shallow water. Another modified version of the KdV equation is the complex mKdV equation ($\epsilon=\pm 1$)
\begin{equation}\label{1.2}
q_{t}+ 6 \epsilon |q|^2 q_x+q_{xxx}=0,
\end{equation}
which is the next member of the nonlinear Schr\"{o}dinger hierarchy (NLS)\cite{a}.
Eq.\eqref{1.2} is completely integrable that can be solved by the inverse scattering transformation method (IST). It possesses all the basic characters of integrable models. The complex mKdV equation has attracted much attention, e.g., the properties of Eq.\eqref{1.2} including interactions of solitary wave, the problem of initial data and structural stability are considered \cite{0.2,0.3}. Existence and stability of solitary wave solution and periodic traveling waves of Eq.\eqref{1.2} are discussed in \cite{0.4,0.5}. Integrable discrete versions of the complex mKdV equation are derived in \cite{0.6,0.7,0.8}. Analytical expressions for the homoclinic orbits and numerical homoclinic instabilities for Eq.\eqref{1.2} are investigated. Conservation laws and exact group invariant solutions are constructed in \cite{0.9}. Breathers and rogue wave solutions of Eq.\eqref{1.2} and the interaction properties of complex solitons are studied in Refs. \cite{0.11,0.12,0.13}.

Very recently, Ablowitz and Musslimani proposed new nonlocal nonlinear integrable equations which include the nonlocal integrable complex mKdV equation \cite{0}
\begin{equation}\label{eq1}
q_{t}+ 6 \epsilon qq^*(-x,-t) q_x+q_{xxx}=0,
\end{equation}
where $q\triangleq q(x,t)$ is a complex function of real variables $x$ and $t$, $*$ denotes complex conjugation, and the nonlocal integrable mKdV equation
\begin{equation}\label{eq1.1}
q_{t}+ 6 \epsilon qq(-x,-t) q_x+q_{xxx}=0.
\end{equation}
Ref.\cite{0.1} gives one soliton solution of the nonlocal complex mKdV \eqref{eq1} ($\epsilon=1$) and nonlocal mKdV equation \eqref{eq1.1} through the IST. The properties of the nonlocal mKdV equation \eqref{eq1.1} including soliton, kink, antikink, complexiton, and rogue-wave have also been discussed in \cite{1,1.1}.

It is interesting to note that the integrable nonlocal NLS equation
\begin{equation}\label{1.1}
i q_t+q_{xx}\pm 2qq^{*}(-x,t)q=0,
\end{equation}
introduced by Ablowitz and Musslimani \cite{0,2} has attracted many researchers due to its special property, e.g., the model \eqref{1.1} admits both bright and dark solitons \cite{3,4}. We have known that the focusing NLS and defocusing NLS are, respectively, gauge equivalent to the Schr\"odinger flow of maps from $R^1$ into $S^2$ in $R^3$ and from $R^1$ into $H^2$ in $R^{2+1}$ \cite{5,6,7,8,9}. It was shown in \cite{10} that the nonlocal NLS equation and its discrete version are gauge equivalent to the Heisenberg-like equation, modified Heisenberg-like equation and their discrete versions respectively. Recently, Gadzhimuradov and Agalarov proved that the nonlocal focusing NLS equation is gauge equivalent to a coupled Landau-Lifshitz equation. The physical and geometrical aspects of this model and their effects on metamagnetic structures are discussed \cite{11}.

In this paper, we focus on the topics of soliton solution and gauge equivalence for the nonlocal complex mKdV equation \eqref{eq1.1}. We show that, under the gauge transformation, the nonlocal complex mKdV equation is gauge equivalent to a spin-like model. From the gauge equivalence, one can see that there exists significant difference between the nonlocal complex mKdV equation and the classical complex mKdV equation. By constructing the DT for nonlocal complex mKdV equation, we derive dark soliton, W-type soliton, M-type soliton and periodic solutions.\\

\section{DT for the nonlocal complex mKdV equation }
In this section, we firstly construct the Darboux transformation \cite{13,14} for nonlocal complex mKdV equation, and then we derive its exact solutions including dark soliton, W-type soliton, M-type soliton and periodic solutions.

Eq.\eqref{eq1} is yielded by the integrability condition of the following spectral equations
\begin{equation}\label{eq2.1}
\varphi_x=M \varphi,\quad \varphi_t=N \varphi,
\end{equation}
where the matrices $M=M(x,t,\lambda)$ and $N=N(x,t,\lambda)$ are given by
\begin{subequations}
\begin{align}
&M=-i\lambda \sigma_3+Q,\\
&N=-4i\lambda^3 \sigma_3+4\lambda^2Q-2i\lambda \sigma_3(Q^2-Q_x)+Q_x Q -Q Q_x-Q_{xx}+2Q^3,
\end{align}
\end{subequations}
with
\begin{align*}
\sigma_3=\left(
\begin{array}{cc}
1 & 0 \\
0 & -1 \\
\end{array}\right),\quad
Q=\left(
\begin{array}{ccc}
0 & q \\
-\epsilon q^*(-x,-t) & 0 \\
\end{array}\right).
\end{align*}
The gauge transformation with nonsingular matrix $P=(p_{jk}(x,t,\lambda))_{2\times2}(j,k=1,2)$
\begin{equation}\label{eq2.2}
\tilde{\varphi}=T \varphi=(\lambda I-P)\varphi
\end{equation}
changes the spectral problem \eqref{eq2.1} into the new one
\begin{equation}\label{eq2.3}
\tilde{\varphi}_x=\tilde{M} \tilde{\varphi},\quad \tilde{\varphi}_t=\tilde{N }\tilde{\varphi}.
\end{equation}
One hopes $\tilde{M},\tilde{N }$ possess the same matrix form as $M,N$, except that the old potential $q$ is substituted by the new potential $\tilde{q}$. It is obvious that Darboux
matrix $T$ satisfies equations
\begin{equation}\label{eq2.4}
T_x=\tilde{M}T-TM,\quad T_t=\tilde{N}T-TN.
\end{equation}
One can obtain the relation between new and old solutions
\begin{equation}\label{eq2.5}
\tilde{q}=q-2ip_{12}, \quad \tilde{q}^{*}(-x,-t)=q^{*}(-x,-t)-2i\epsilon p_{21},
\end{equation}
with a constraint
\begin{equation}\label{eq2.6}
p_{12}=-\epsilon p^{*}_{21}(-x,-t).
\end{equation}
By setting
\begin{equation}\label{eq2.7}
P=H\Lambda H^{-1},
\end{equation}
with
\begin{align*}
H=\left(
\begin{array}{cc}
f_1 & g_1 \\
f_2 & g_2  \\
\end{array}\right),\quad
\Lambda=\left(
\begin{array}{cc}
\lambda_1 & 0 \\
0 & \lambda_2  \\
\end{array}\right),
\end{align*}
where $(f_1,f_2)^T=(f_1(x,t),f_2(x,t))^T$ is a solution to Eq.\eqref{eq2.1} with $\lambda=\lambda_1$ and
$(g_1,g_2)^T=(\epsilon f^{*}_2(-x,-t),f^{*}_1(-x,-t))^T$ is the solution when $\lambda=-\lambda^{*}_1\triangleq \lambda_2$,
we can obtain the explicit expression of $P$,
\begin{align}\label{eq2.8}
P=\frac{1}{\Delta}\left(
\begin{array}{cc}
\lambda_1 f_1f^{*}_1(-x,-t)+\epsilon\lambda^{*}_1f_2f^{*}_2(-x,-t) & -\epsilon(\lambda_1+\lambda^{*}_1)f_1f^{*}_2(-x,-t) \\
(\lambda_1+\lambda^{*}_1)f_2 f^{*}_1(-x,-t)& -\epsilon\lambda_1 f_2f^{*}_2(-x,-t)-\lambda^{*}_1f_1f^{*}_1(-x,-t)  \\
\end{array}\right),
\end{align}
with $\Delta=f_1f^{*}_1(-x,-t)-\epsilon f_2f^{*}_2(-x,-t)$. Hence, the new solution is written as
\begin{equation}\label{eq2.9}
\tilde{q}=q+\frac{2i\epsilon}{\Delta}(\lambda_1+\lambda^{*}_1)f_1f^{*}_2(-x,-t).
\end{equation}
The $n$-fold DT in the form of a determinant is the same one as the Ref.\cite{15}
\begin{equation}\label{eq2.10}
\tilde{q}[n]=q-2i\frac{P_{2n}}{W_{2n}},
\end{equation}
where
\begin{equation*}
P_{2n}=\left|
\begin{array}{ccccccc}
f_1 & f_2 & \lambda_1 f_1 & \lambda_1 f_2 & \cdot\cdot\cdot & \lambda_1^{n-1} f_1 & \lambda_1^n f_1 \\
g_1 & g_2 & \lambda_2 g_1 & \lambda_2 g_2 & \cdot\cdot\cdot & \lambda_2^{n-1} g_1 & \lambda_2^n g_1\\
f_3 & f_4 & \lambda_3 f_3 & \lambda_3 f_4 & \cdot\cdot\cdot & \lambda_3^{n-1} f_3 & \lambda_3^n f_3 \\
g_3 & g_4 & \lambda_4 g_3 & \lambda_4 g_4 & \cdot\cdot\cdot & \lambda_4^{n-1} g_3 & \lambda_4^n g_3 \\
\vdots & \vdots & \vdots  & \vdots  & \ddots & \vdots  & \vdots  \\
g_{2n-1} & g_{2n} & \lambda_{2n} g_{2n-1} & \lambda_{2n} g_{2n} & \cdot\cdot\cdot & \lambda_{2n}^{n-1} g_{2n-1} & \lambda_{2n}^n g_{2n-1} \\
\end{array}\right|,
\end{equation*}
\begin{equation*}
W_{2n}=\left|
\begin{array}{ccccccc}
f_1 & f_2 & \lambda_1 f_1 & \lambda_1 f_2 & \cdot\cdot\cdot & \lambda_1^{n-1} f_1 & \lambda_1^{n-1} f_2 \\
g_1 & g_2 & \lambda_2 g_1 & \lambda_2 g_2 & \cdot\cdot\cdot & \lambda_2^{n-1} g_1 & \lambda_2^{n-1} g_2\\
f_3 & f_4 & \lambda_3 f_3 & \lambda_3 f_4 & \cdot\cdot\cdot & \lambda_3^{n-1} f_3 & \lambda_3^{n-1} f_4 \\
g_3 & g_4 & \lambda_4 g_3 & \lambda_4 g_4 & \cdot\cdot\cdot & \lambda_4^{n-1} g_3 & \lambda_4^{n-1} g_4 \\
\vdots & \vdots & \vdots  & \vdots  & \ddots & \vdots  & \vdots  \\
g_{2n-1} & g_{2n} & \lambda_{2n} g_{2n-1} & \lambda_{2n} g_{2n} & \cdot\cdot\cdot & \lambda_{2n}^{n-1} g_{2n-1} & \lambda_{2n}^{n-1} g_{2n} \\
\end{array}\right|.
\end{equation*}
It should be remarked that $P_{2n}$ admits the following rule:
\begin{equation*}
P_2=\left|
\begin{array}{cc}
f_1 & \lambda_1 f_1 \\
g_1 & \lambda_2 g_1\\
\end{array}\right|, \quad
P_4=\left|
\begin{array}{cccc}
f_1 & f_2 & \lambda_1 f_1  & \lambda_1^2 f_1 \\
g_1 & g_2 & \lambda_2 g_1  & \lambda_2^2 g_1\\
f_3 & f_4 & \lambda_3 f_3  & \lambda_3^2 f_3 \\
g_3 & g_4 & \lambda_4 g_3  & \lambda_4^2 g_3  \\
\end{array}\right|, ......
\end{equation*}
Next, we will construct exact solutions of nonlocal complex mKdV equation \eqref{eq1} by using the obtained DT. Let us consider the nonlocal complex mKdV$^-$($\epsilon=-1$).\\
\textbf{Case 1.} For zero seed solution $q=0$, the new solution with $\lambda_1=a+ib(a\neq 0)$ is
\begin{align}\label{eq4.11}
\tilde{q}=-2iae^{2b(x+4(3a^2-b^2)t)}\sec(2a(x+4(a^2-3b^2)t)),
\end{align}
which has singularities at $\{(x,t)|x+4(a^2-3b^2)t=k\pi+\frac{\pi}{2}, k\in Z\}$.\\
\textbf{Case 2.} For nonzero seed solution
\begin{eqnarray}\label{eq4.12}
q=\rho e^{k(x-(k^2-6|\rho|^2)t)}\triangleq \rho e^{\Omega},
\end{eqnarray}
where $k$ is a real parameter and $\rho\neq0$ is a complex one. Solving Eq.\eqref{eq2.1} gives
\begin{align}\label{eq4.13}
&f_1=e^{\frac{k}{2}(x-(k^2-6|\rho|^2)t)}\left(c_1 e^{\frac{\xi}{2}(x+\eta t)}+c_2 e^{-\frac{\xi}{2}(x+\eta t)}\right),\\ \label{eq4.14}
&f_2=\frac{1}{2\rho}e^{-\frac{k}{2}(x-(k^2-6|\rho|^2)t)}\left(c_1(k+\xi+2i\lambda) e^{\frac{\xi}{2}(x+\eta t)}+c_2(k-\xi+2i\lambda) e^{-\frac{\xi}{2}(x+\eta t)}\right),
\end{align}
with $\xi=\sqrt{(k+2i\lambda_1)^2+4|\rho|^2}$ and $ \eta=4\lambda_1^2+2i \lambda_1 k-k^2+2|\rho|^2$, where $c_1$ and $c_2$ are nonzero complex parameters.

Let $k=2b$, then $\xi=2\sqrt{|\rho|^2-a^2},\eta=4a^2-12b^2+2|\rho|^2+12 i ab\triangleq \eta_1+i \eta_2$.

\begin{figure*}[htb]
\centering
\subfigure[$\textrm{dark soliton}$] {\includegraphics[width=0.3\textwidth]{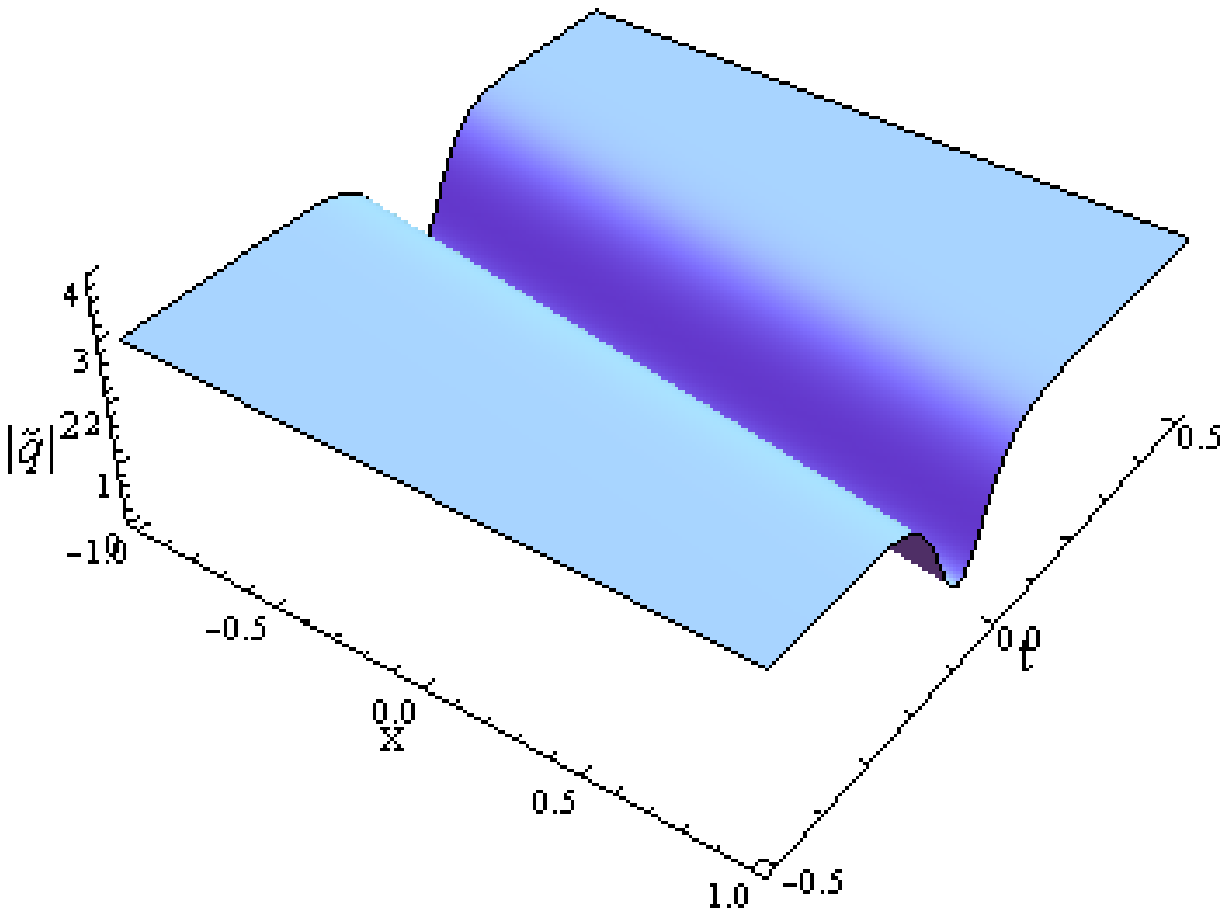}} \quad
\subfigure[$\textrm{W-type soliton}$] {\includegraphics[width=0.3\textwidth]{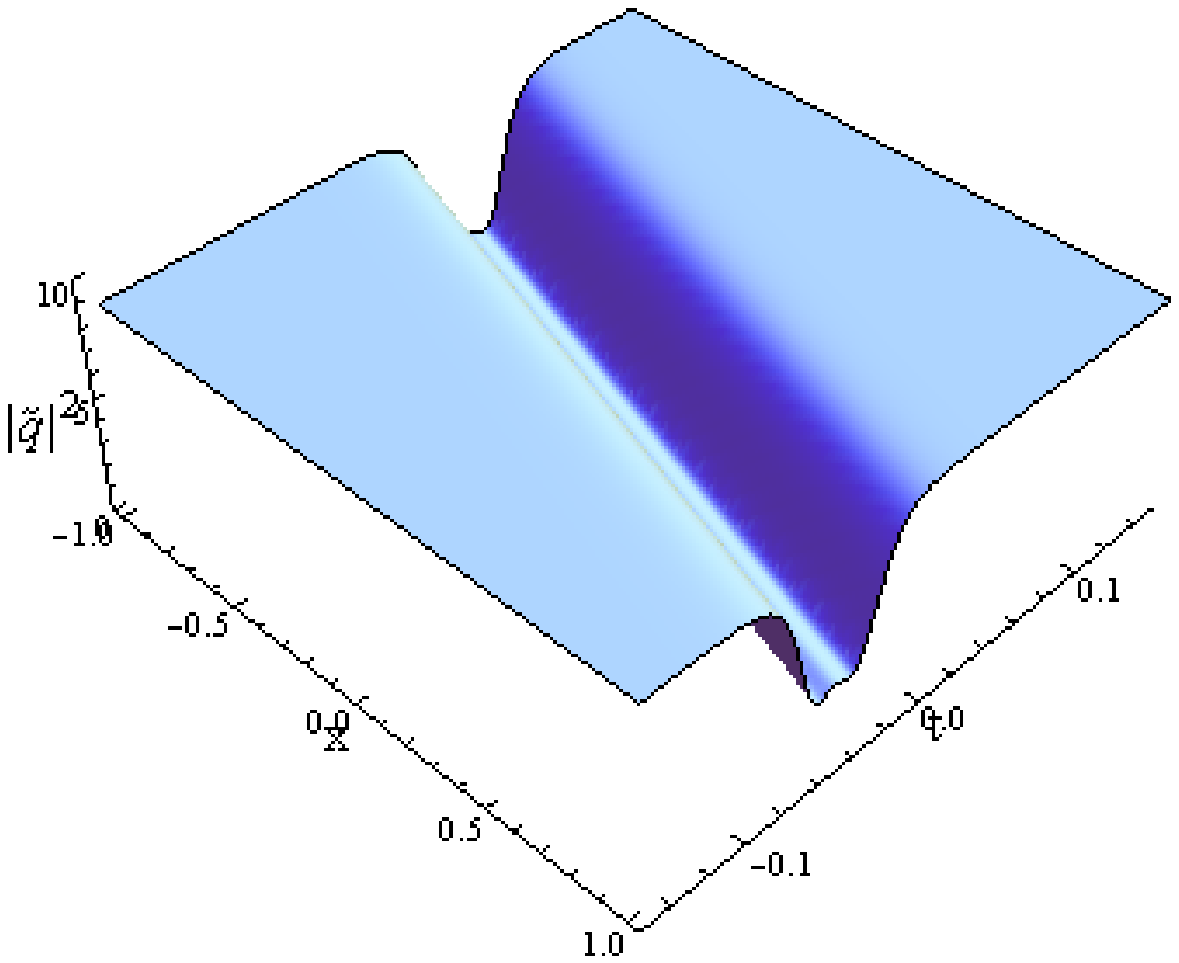}} \quad
\subfigure[$\textrm{M-type soliton}$] {\includegraphics[width=0.3\textwidth]{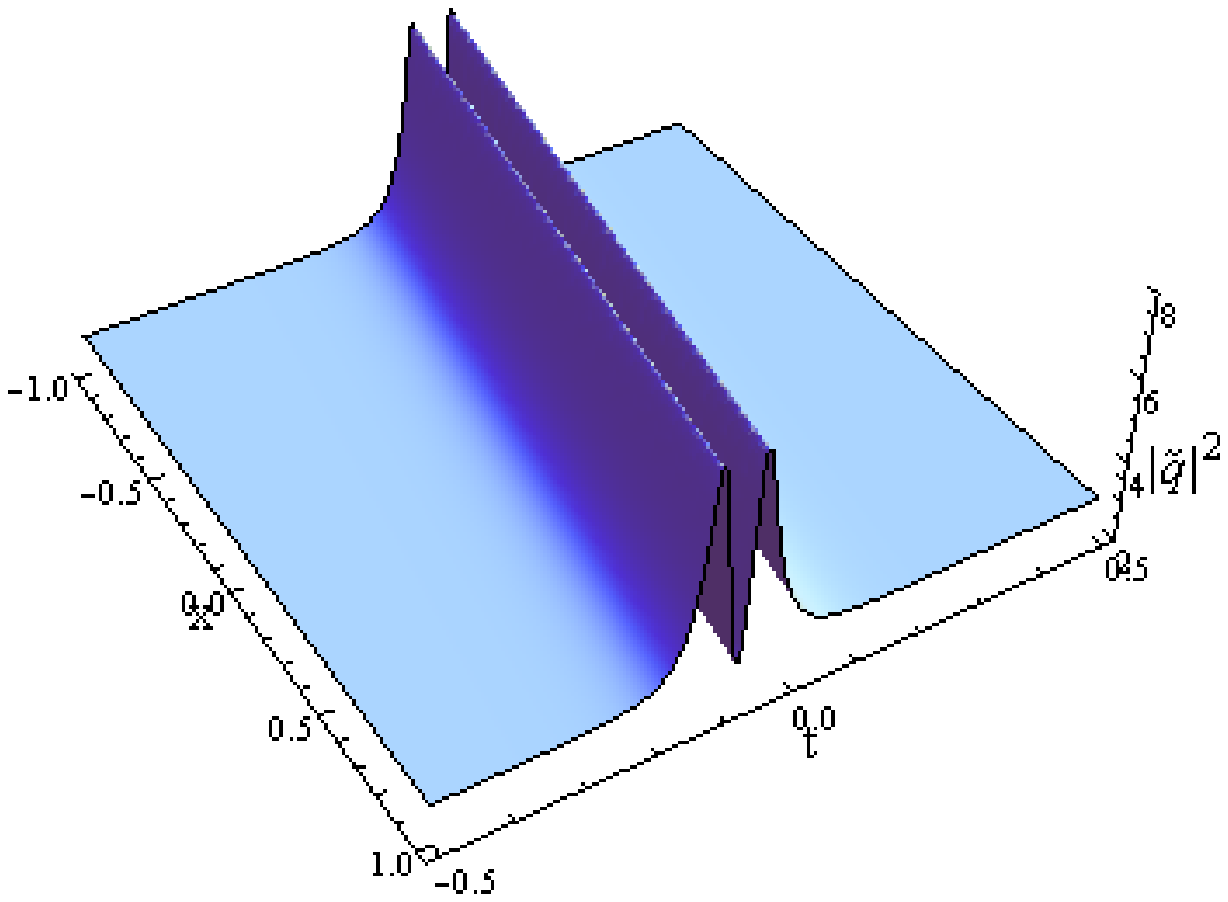}}
\caption{\small{(a): dark soliton with $a=\sqrt{2},b=0,\rho=\sqrt{3}i,\gamma=1$; (b): W-type soliton $a=1,b=0,\rho=3,\gamma=1$;
(c): M-type soliton with $a=\sqrt{2},b=0,\rho=\sqrt{3}i,\gamma=-1$.}}
\label{Fig1}
\end{figure*}

\begin{figure*}[htb]
\centering
\subfigure[$$] {\includegraphics[width=0.32\textwidth]{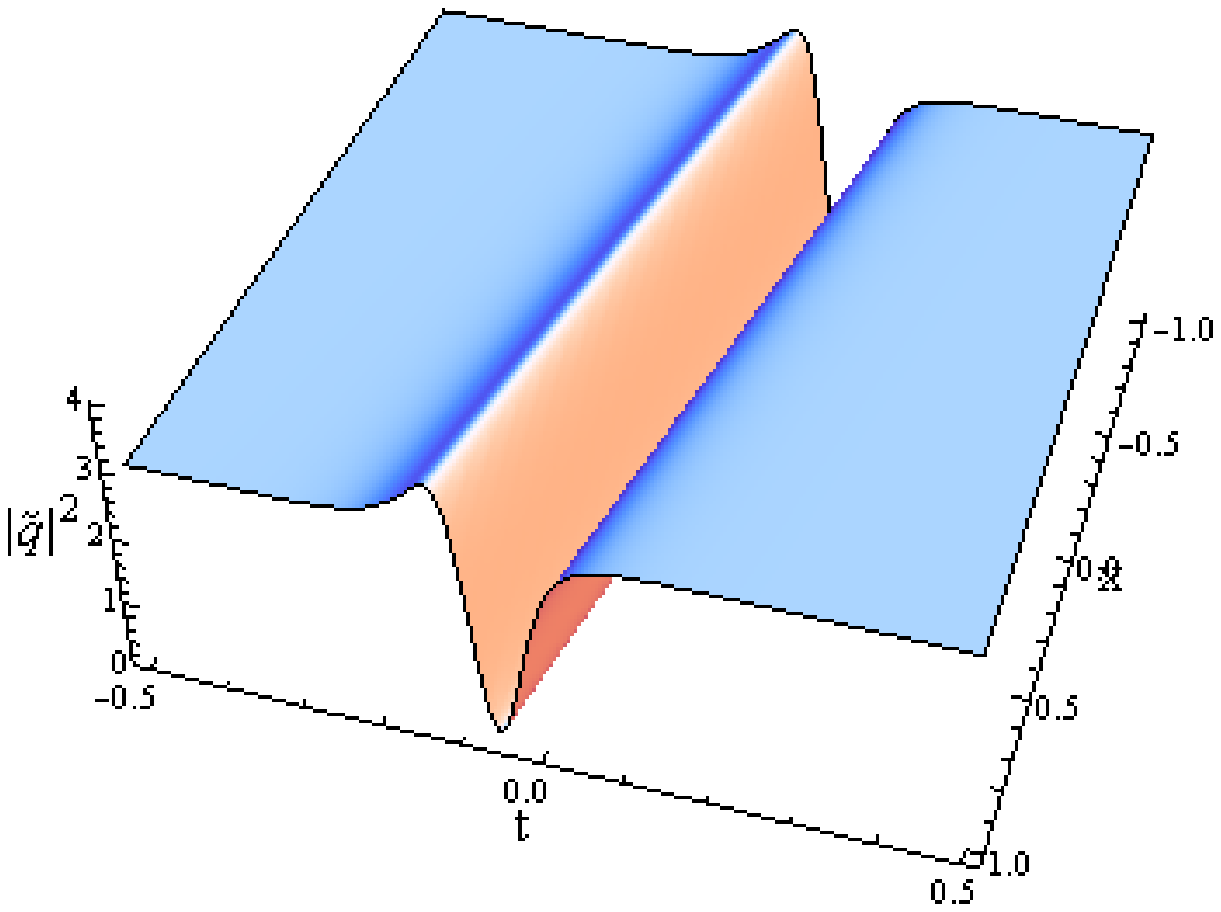}} \qquad
\subfigure[$$] {\includegraphics[width=0.32\textwidth]{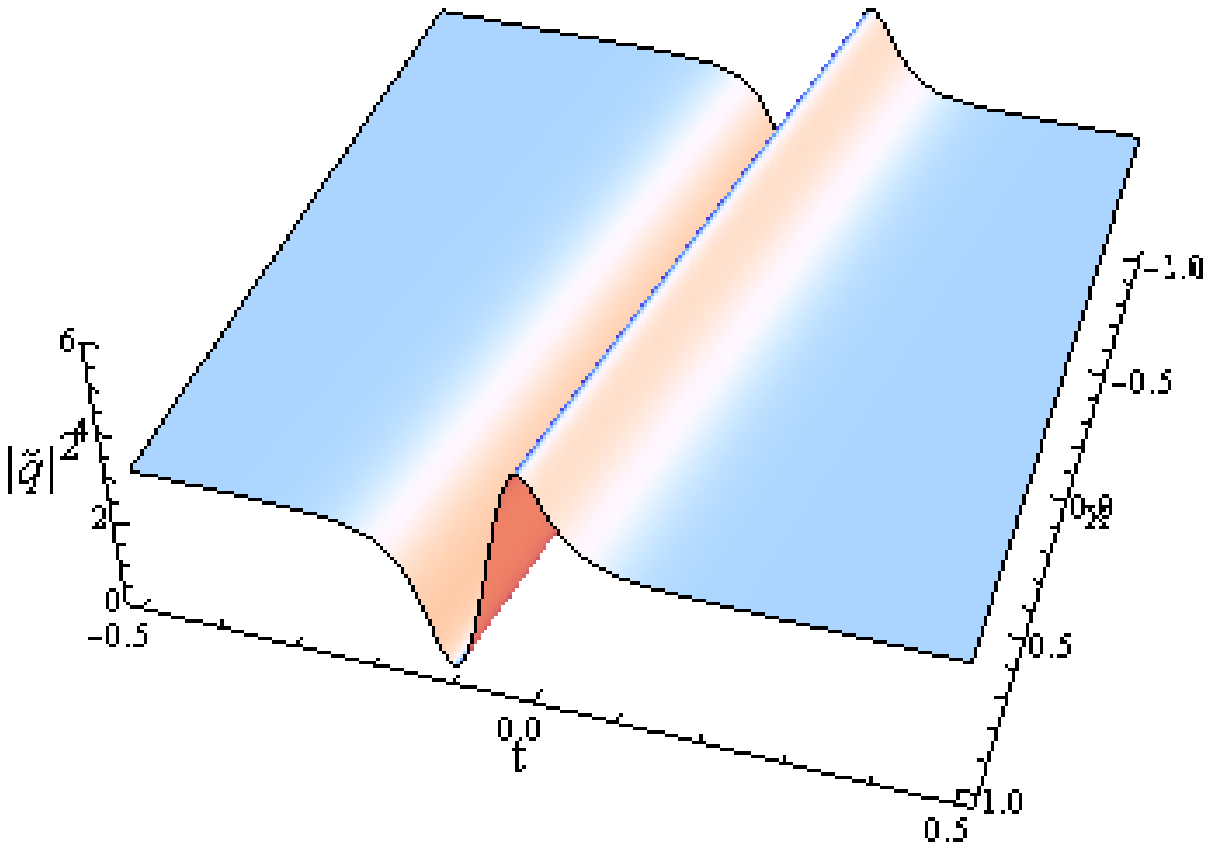}}
\caption{\small{Solution (a): $a=\sqrt{2},b=0,\rho=\sqrt{3}i,\gamma=-1-i$; (b): $a=\sqrt{2},b=0,\rho=\sqrt{3}i,\gamma=1+2i$.}}
\label{Fig2}
\end{figure*}

$\textbf{(i)}$ If $|a|<|\rho|$, we have $\xi=2\sqrt{|\rho|^2-a^2}\triangleq 2s$. The eigenfunctions
\eqref{eq4.13} and \eqref{eq4.14} are
\begin{align}\label{eq4.15}
&f_1=e^{b(x-2(2b^2-3|\rho|^2)t)}\left(c_1 e^{s(x+\eta_1 t)+i s \eta_2 t}+c_2 e^{-s(x+\eta_1 t)-is \eta_2 t}\right),\\ \label{eq4.16}
&f_2=e^{-b(x-2(2b^2-3|\rho|^2)t)}\left(\frac{c_1(s+i a)}{\rho} e^{s(x+\eta_1 t)+i s \eta_2 t}-\frac{c_2(s-ia)}{\rho} e^{-s(x+\eta_1 t)-is \eta_2 t}\right).
\end{align}
So new solution with $\gamma=c_2/c_1$ is
\begin{eqnarray}\label{eq4.17}
\tilde{q}=\rho e^{\Omega}(1-2a  \frac{(a+is)e^{2is \eta_2 t}+\gamma^*(a-is)e^{2s(x+\eta_1 t)}+\gamma(a+is)e^{-2s(x+\eta_1 t)}+|\gamma|^2(a-is)e^{-2is \eta_2 t}}{|\rho|^2e^{2is \eta_2 t}+a\gamma^*(a-is)e^{2s(x+\eta_1 t)}+a\gamma(a+is) e^{-2s(x+\eta_1 t)}+|\rho|^2|\gamma|^2e^{-2is \eta_2 t}}),
\end{eqnarray}
which can be simplified as
\begin{eqnarray}\label{eq4.18}
\tilde{q}=-\rho\frac{2a\gamma(a\cosh(2s(x+\eta_1t))+is\sinh(2s(x+\eta_1t)))+(2a^2-|\rho|^2)(1+\gamma^2)+ 2ias(1-\gamma^2)}{2a\gamma(a\cosh(2s(x+\eta_1t))-is\sinh(2s(x+\eta_1t)))+|\rho|^2(1+\gamma^2)},
\end{eqnarray}
as $b=0$, i.e., $\lambda_1\in R$, and $\textrm{Im}\gamma=0$.
The norm of \eqref{eq4.18} is given by
\begin{eqnarray}\label{eq4.19}
|\tilde{q}|^2=\frac{4|\rho|^2a^2s^2[-1+\gamma(\sinh(2s(x+\eta_1t))+\gamma)]^2+(2a^2\gamma\cosh(2s(x+\eta_1t))+(2a^2-|\rho|^2)(1+\gamma^2))^2}
{4a^2s^2\gamma^2\sinh^2(2s(x+\eta_1t))+(2a^2\gamma\cosh(2s(x+\eta_1t))+|\rho|^2(1+\gamma^2))^2}.
\end{eqnarray}
Next we analyze its dynamic properties.  From \eqref{eq4.19}, we can see $|\tilde{q}|^2\rightarrow |\rho|^2$ as $t\rightarrow \pm \infty$.  For convenience, let $e^{-2s(x+\eta_1t)}=X>0$ in \eqref{eq4.19}, then $d|\tilde{q}|^2/dX=0$ gives
\begin{eqnarray}\label{eq4.20}
f'=(1+\gamma X)(\alpha_1 X^5+\alpha_2X^4+\alpha_3X^3+\alpha_4X^2+\alpha_5X+\alpha_6)=0,
\end{eqnarray}
where
\begin{equation*}\begin{aligned}
&\alpha_6=a^2|\rho|^2\gamma^2,\quad \alpha_1=-\gamma \alpha_6,\quad \alpha_2=-3\alpha_6, \quad \alpha_5=3\gamma\alpha_6,\\
&\alpha_3=\gamma\left(-4a^2|\rho|^2+|\rho|^4+2\gamma^2(2a^4-3 a^2|\rho|^2+|\rho|^4)+|\rho|^4\gamma^4\right)\\
&\alpha_4=-|\rho|^4-2\gamma^2(2a^4-3a^2|\rho|^2+|\rho|^4)-|\rho|^2\gamma^4(|\rho|^2-4a^2).
\end{aligned}\end{equation*}
The shape of solution depends on the parameters $a, \rho$ and $\gamma$ which determine the number of positive real root of \eqref{eq4.20}. Here we list  several cases.\\
$\bullet$ A dark soliton solution is given in Fig.\ref{Fig1}(a) when $a=\sqrt{2},b=0,\rho=\sqrt{3}i,\gamma=1$. The condition $f'(1)=0$ and $f''(1)>0$ indicates $f_{min}(X)=f(1)$.  \\
$\bullet$ Fig.\ref{Fig1}(b) describes the W-type solution with $a=1,b=0,\rho=3,\gamma=1$, where extreme values are lower than background plane $|\tilde{q}|^2=9$. \\
$\bullet$ The M-type soliton is described in Fig.\ref{Fig1}(c) where we take $a=\sqrt{2},b=0,\rho=\sqrt{3}i,\gamma=-1$.\\
However, as $b=0, \textrm{Im}\gamma\neq 0$, the shape of soliton depends on $\gamma$ with other parameters fixed. For instance, when $a=\sqrt{2},b=0,\rho=\sqrt{3}i$, the soliton is depicted in Fig.\ref{Fig2} with $\gamma=-1-i$ and $\gamma=1+2i$. We can see that
maximum value of $|\tilde{q}|^2$ is higher than the background plane, but minimum value is lower than the background plane.

\begin{figure*}[!t]
\centering
\subfigure[] {\includegraphics[width=0.3\textwidth]{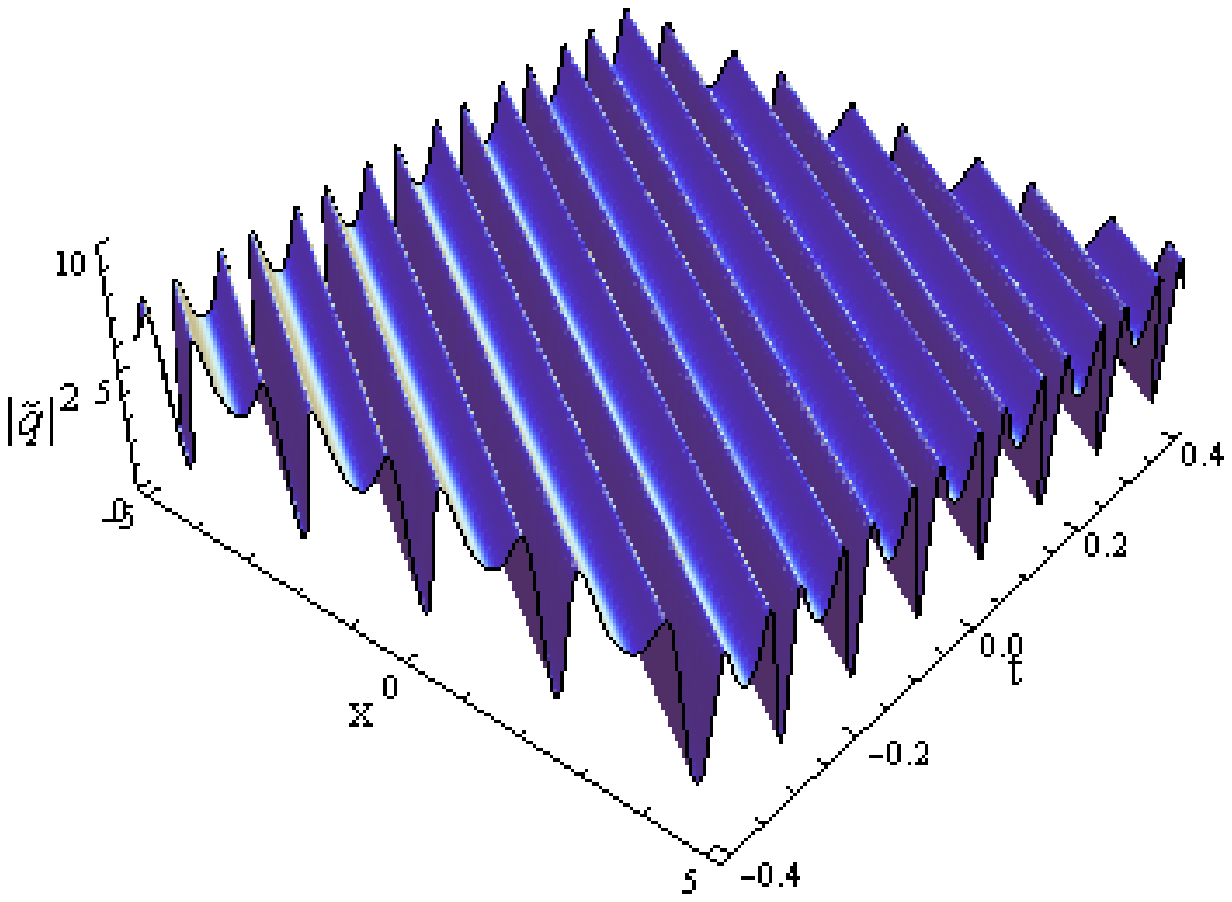}} \quad
\subfigure[$u=\tilde{q}\tilde{q}^{*}(-x,-t)$] {\includegraphics[width=0.3\textwidth]{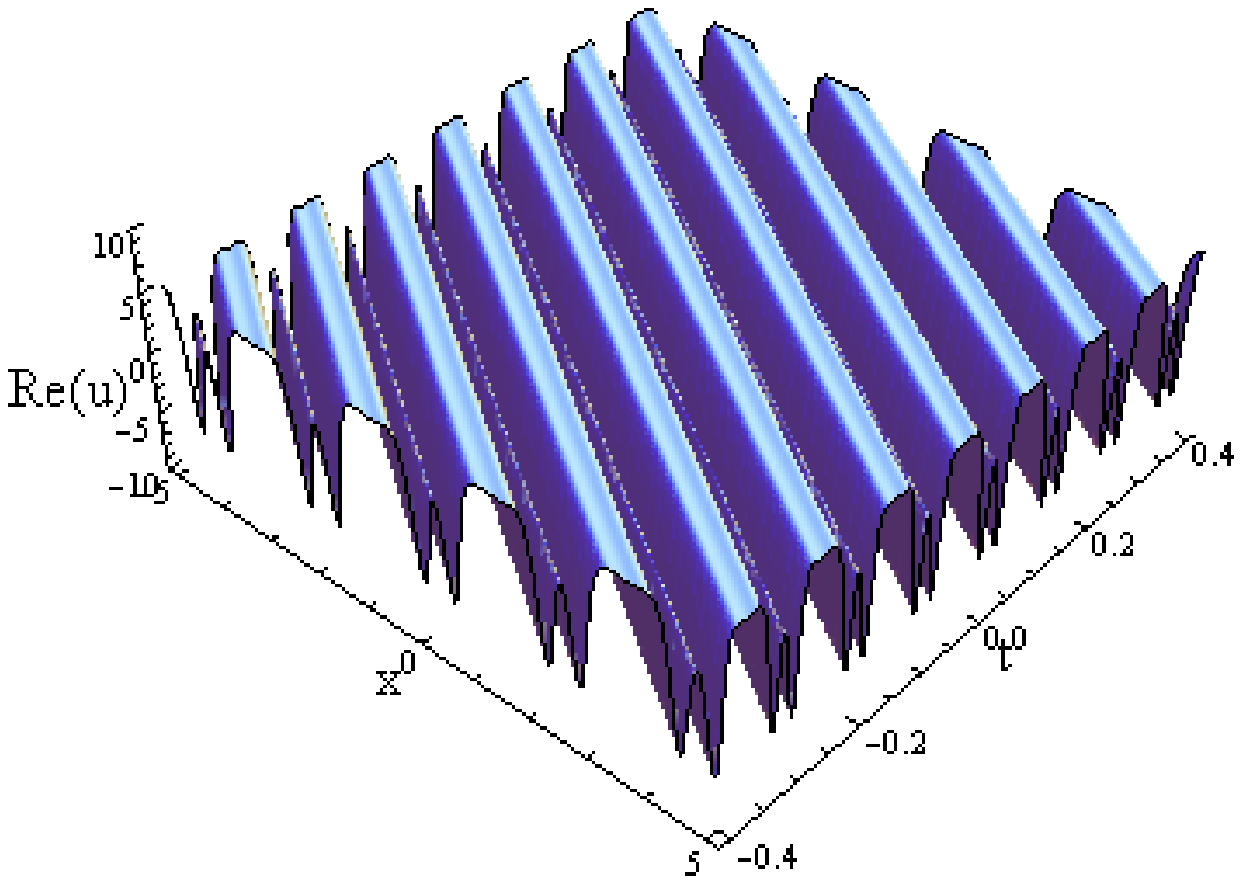}} \quad
\subfigure[$u=\tilde{q}\tilde{q}^{*}(-x,-t)$] {\includegraphics[width=0.3\textwidth]{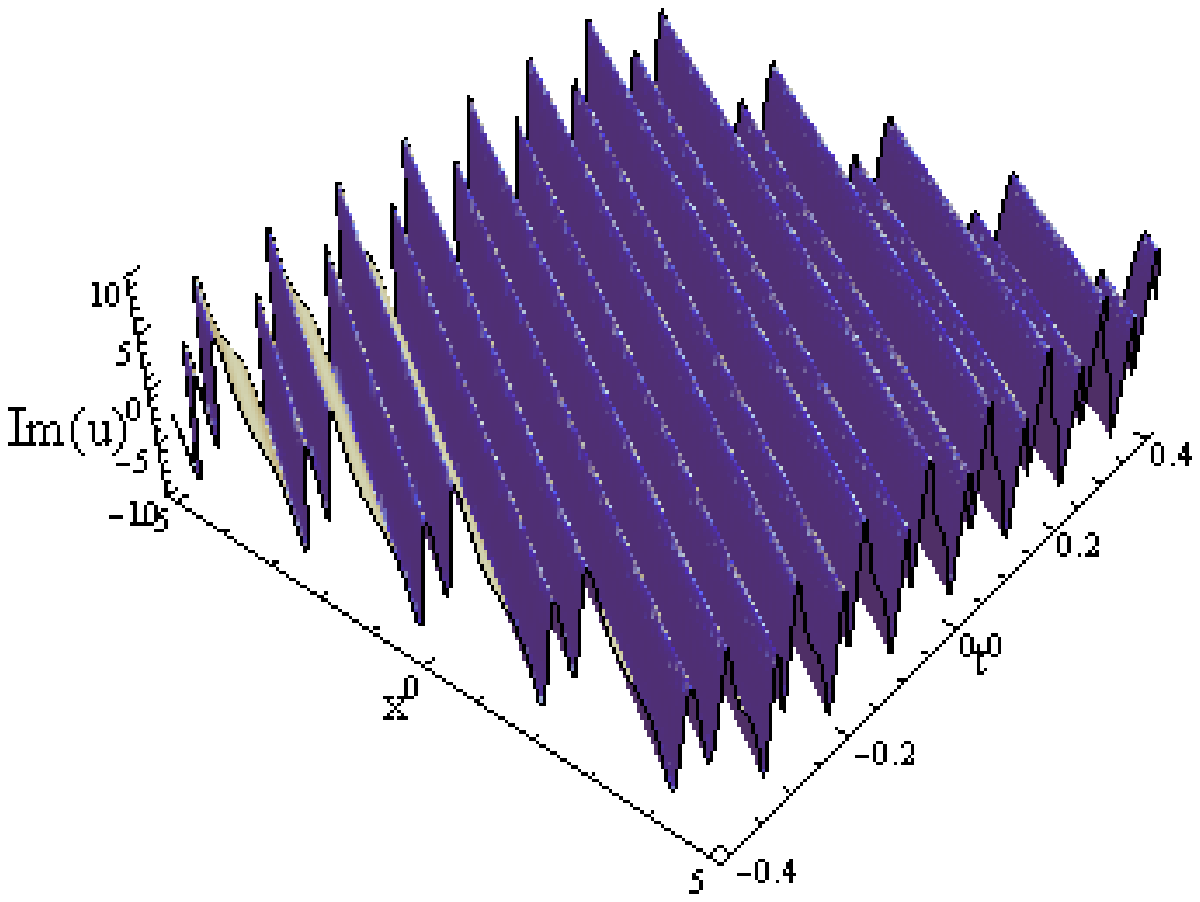}} \\
\subfigure[] {\includegraphics[width=0.28\textwidth]{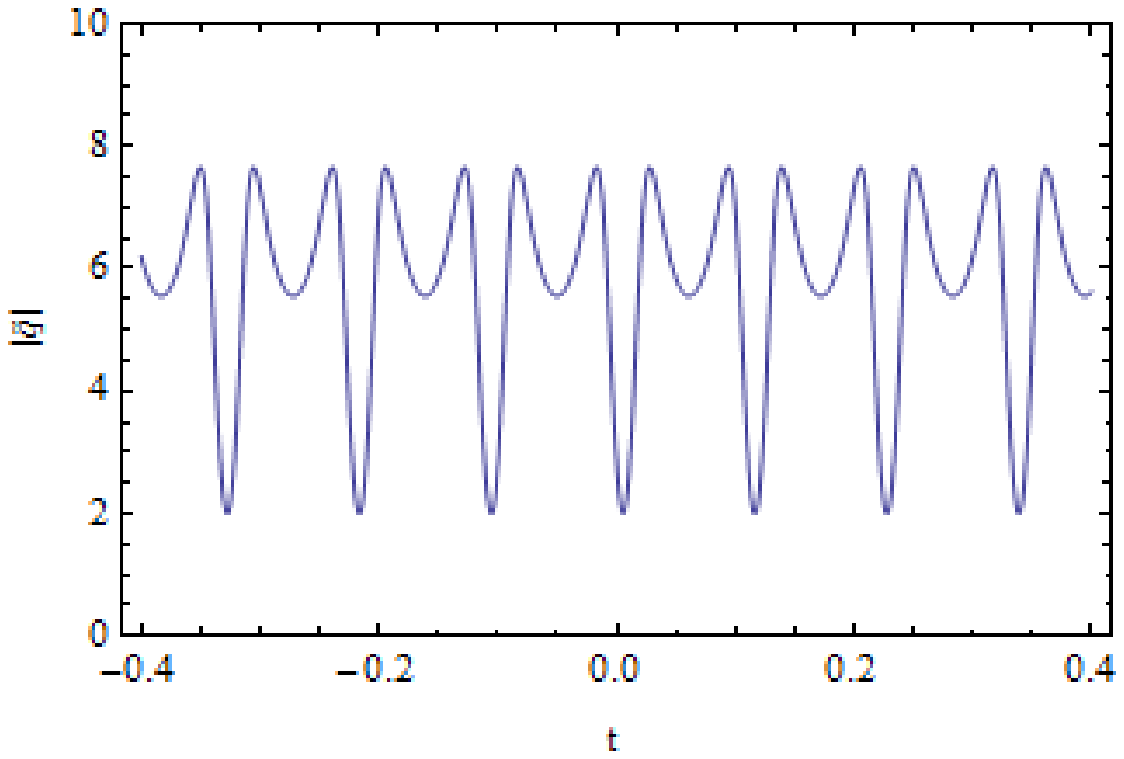}} \quad
\subfigure[] {\includegraphics[width=0.28\textwidth]{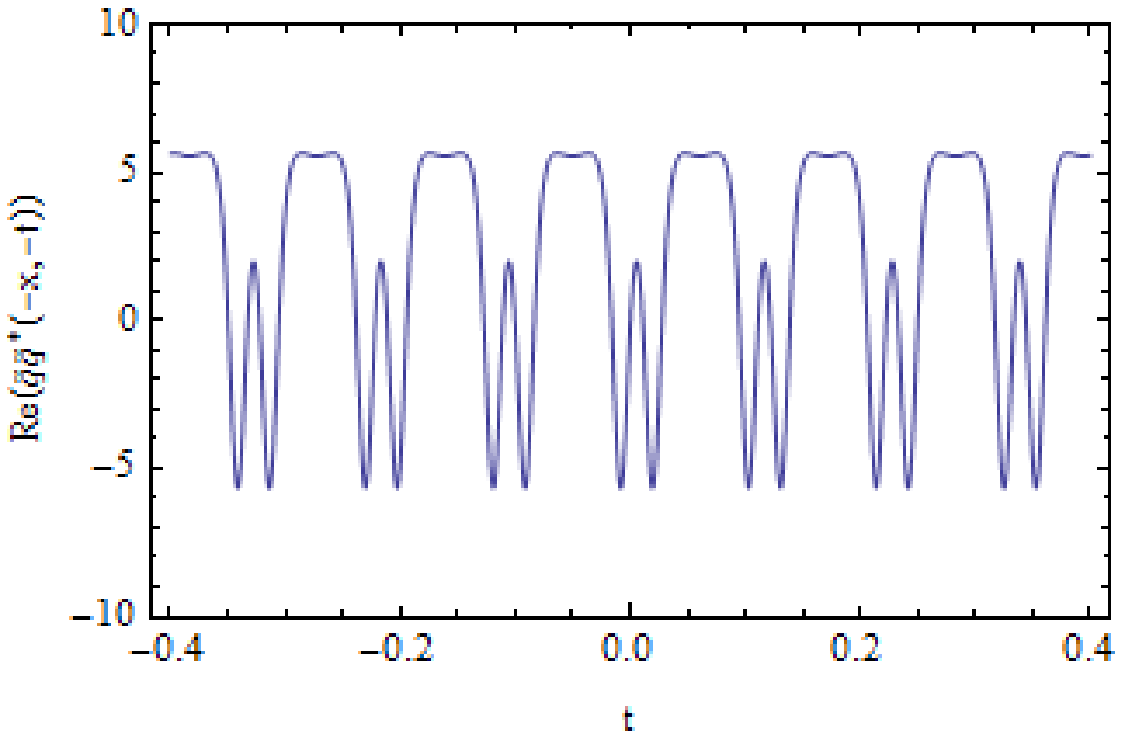}} \quad
\subfigure[] {\includegraphics[width=0.28\textwidth]{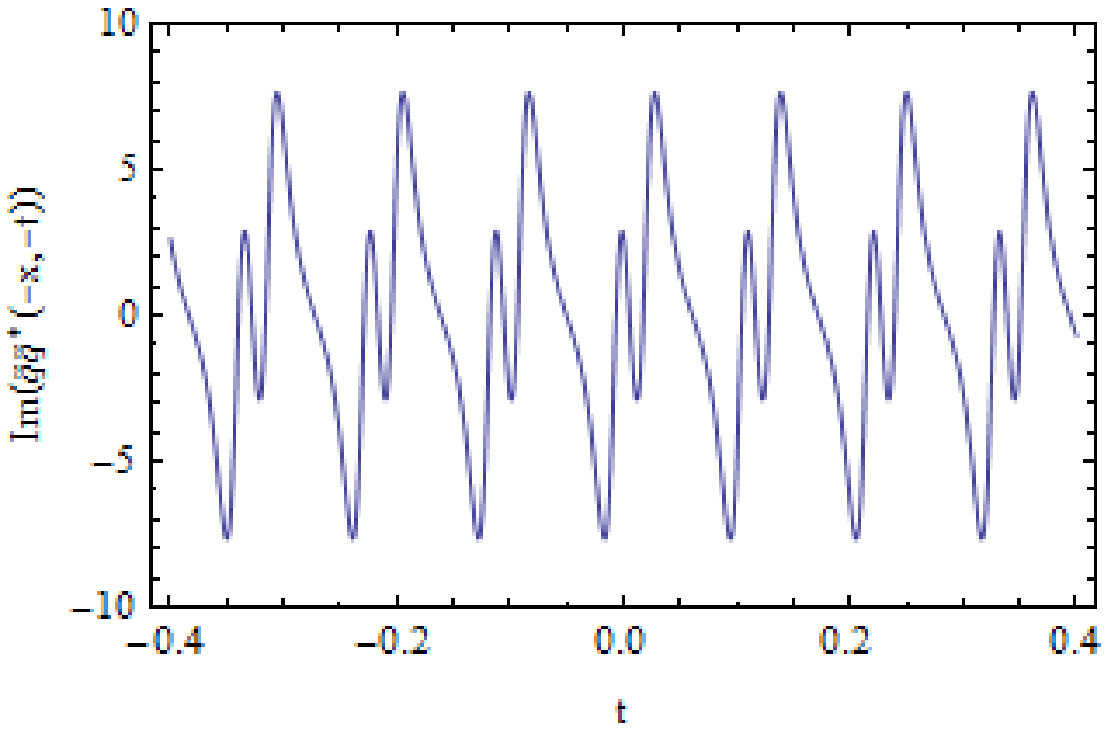}} \\
\subfigure[] {\includegraphics[width=0.28\textwidth]{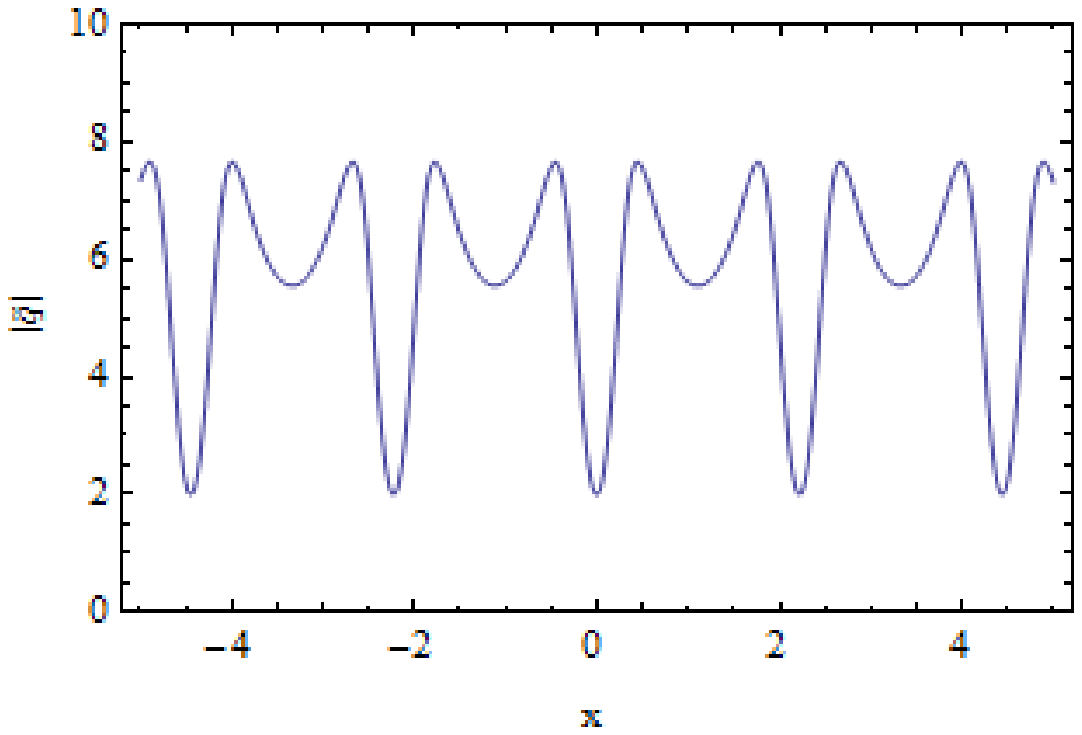}} \quad
\subfigure[] {\includegraphics[width=0.28\textwidth]{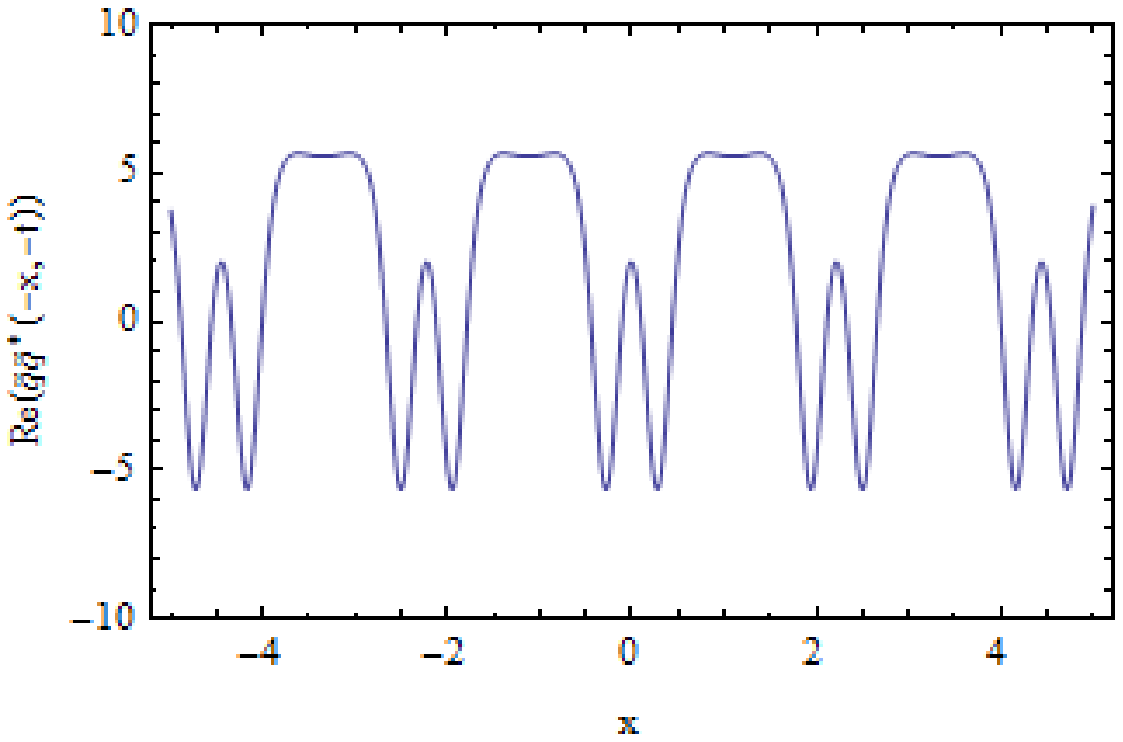}} \quad
\subfigure[] {\includegraphics[width=0.28\textwidth]{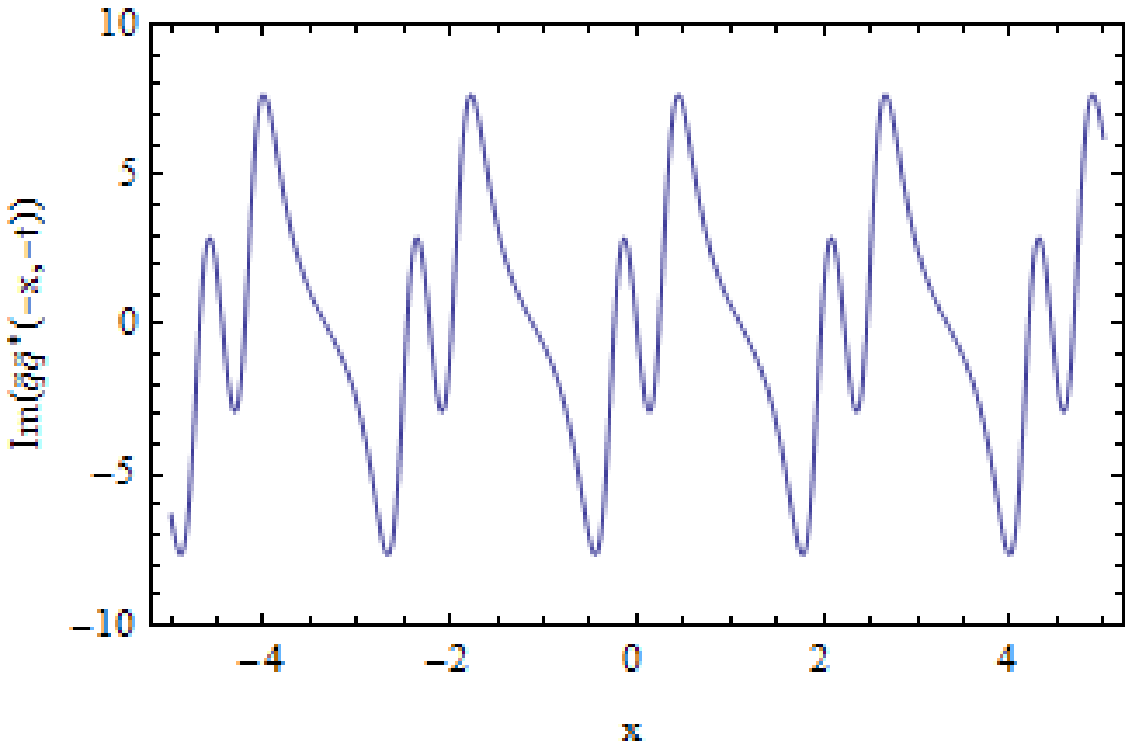}}
\caption{\small{The periodic solution \eqref{eq4.18} with $T_{space}=\frac{\sqrt{2}\pi}{2}\approx 2.221$, $T_{time}=\frac{\sqrt{2}\pi}{40}\approx 0.111$ as parameters $a=2,b=0,\gamma=1,\rho=\sqrt{2}$.  (a-c) 3D plot: $|\tilde{q}|$, $\textrm{Re}(\tilde{q}\tilde{q}^*(-x,-t))$ and $\textrm{Im}(\tilde{q}\tilde{q}^*(-x,-t))$; (d-f) Time period at $x=1$; (g-i) Space period at $t=0.5$. }}
\label{Fig3}
\end{figure*}

$\textbf{(ii)}$ If $|a|>|\rho|$, this means $\xi=2i\sqrt{a^2-|\rho|^2}\triangleq 2il$. Under the conditions: $b=0$, $a+l\neq (a-l)|\gamma|^2$, and $a(a+l)+a(a-l)|\gamma|^2\neq \pm|\rho|^2(\gamma+\gamma^*)$, we get a periodic solution
\begin{eqnarray}\label{eq4.21}
\tilde{q}=\rho \left(1-2a \frac{(a+l)e^{2il(x+\eta_1 t)}+|\gamma|^2(a-l)e^{-2il(x+\eta_1 t)}+\gamma(a+l)+\gamma^*(a-l)}{a(a+l) e^{2il(x+\eta_1 t)}+a|\gamma|^2(a-l) e^{-2il(x+\eta_1 t)}+|\rho|^2(\gamma+\gamma^*)}\right),
\end{eqnarray}
with the period $T_{space}=\frac{\pi}{l}$ and $T_{time}=\frac{\pi}{l\eta_1}$ in space and time, respectively. Fig.\ref{Fig3} depicts the shape of $|\tilde{q}|$, $\textrm{Re}(\tilde{q}\tilde{q}^*(-x,-t))$ and $\textrm{Im}(\tilde{q}\tilde{q}^*(-x,-t))$ with $a=2,b=0,\gamma=1,\rho=\sqrt{2}$.

\section{Gauge equivalence of the nonlocal complex mKdV$^-$ equation }
In this section, we show that, under the gauge transformation, the nonlocal complex mKdV$^-$ equation is gauge equivalent to a coupled spin equation. Then, we construct solution of the coupled spin equation based on the solution of the nonlocal complex mKdV$^-$ equation.

Let us consider the nonlocal complex mKdV$^-$ equation. Under the gauge transformation:
\begin{equation}\label{gt}
\tilde{\varphi} =G^{-1}\varphi, \quad \tilde{M}=G^{-1}M G-G^{-1}G_{x}, \quad \tilde{N}=G^{-1}N G-G^{-1}G_{t},
\end{equation}
where $G$ is a solution of the system \eqref{eq2.1} for $\lambda=0$, i.e.,
\begin{equation}\label{eq3}
G_{x}=M(0)G, \quad G_{t}=N(0)G.
\end{equation}
\begin{subequations}
Define $S=-iG^{-1}\sigma_3G$. We have
\begin{align}\label{eq4.3}
&\tilde{M}=\lambda S, \\ \label{eq4.4}
&\tilde{N}=4\lambda^3S-2\lambda^2 SS_x+\lambda(\frac{3}{2}SS_x^2-S_{xx}),
\end{align}
\end{subequations}
where three important identities have been used,
\begin{equation}
SS_x=-2G^{-1}QG, \quad SS_x^2=-4iG^{-1}\sigma_3Q^2G,\quad S_{xx}-SS_x^2=-2iG^{-1}\sigma_3Q_xG.
\end{equation}
Consider the spectral equations $\tilde{\varphi}_x=\tilde{M}\tilde{\varphi}, \quad \tilde{\varphi}_t=\tilde{N}\tilde{\varphi}$. The zero curvature condition $\tilde{M}_t-\tilde{N}_x+[\tilde{M},\tilde{N}]=0$ yields an integrable spin-like equation
\begin{equation}\label{eq4.5}
S_t+S_{xxx}-\frac{3}{2}\left(S_x^3+SS_{xx}S_{x}+SS_{x}S_{xx}\right)=0.
\end{equation}

Next, we construct the exact solution $S$ of equation \eqref{eq4.5} through the solution $q$ of mKdV$^-$ equation with the help of the gauge transformation \eqref{gt}. From the structure of the matrices $M(0)$ and $N(0)$, we can observe that $G$ in \eqref{eq3} has the form
\begin{equation}\label{eq4.6}
G=\left(
\begin{array}{cc}
f & -g^{*}(-x,-t) \\
g & f^{*}(-x,-t) \\
\end{array}\right),
\end{equation}
Hence the matrix $S$ has the form
\begin{equation}\label{eq4.7}
S=-iG^{-1}\sigma_3 G=\frac{i}{\Delta}\left(
\begin{array}{cc}
gg^{*}(-x,-t)-ff^{*}(-x,-t) & 2f^{*}(-x,-t)g^{*}(-x,-t) \\
2fg & ff^{*}(-x,-t)-gg^{*}(-x,-t) \\
\end{array}\right),
\end{equation}
where $\Delta=ff^{*}(-x,-t)+gg^{*}(-x,-t)$.\\
For the given solution $q^{[1]}$ described by equation \eqref{eq4.11}, solving the Eq. \eqref{eq3} yields
\begin{eqnarray}\label{eq4.8}
G=\left(
\begin{array}{cc}
\frac{a e^{2b(x+4(3a^2-b^2)t)}\sec(2a(x+4(a^2-3b^2)t))}{\sqrt{a^2+b^2}} & i \frac{b- a\tan(2a(x+4(a^2-3b^2)t)}{\sqrt{a^2+b^2}} \\\\
i \frac{b+a\tan(2a(x+4(a^2-3b^2)t)}{\sqrt{a^2+b^2}} & \frac{a e^{-2b(x+4(3a^2-b^2)t)}\sec(2a(x+4(a^2-3b^2)t))}{\sqrt{a^2+b^2}} \\
\end{array}\right).
\end{eqnarray}
We thus obtain the exact solution $S$ of Eq. \eqref{eq4.5}
\begin{equation*}
S=\left(
\begin{array}{cc}
S_{11} & S_{12} \\
S_{21} & -S_{11} \\
\end{array}\right),
\end{equation*}
where
\begin{align*}\nonumber
&S_{11}=i\frac{[-3a^2+b^2+(a^2+b^2)\cos(4a(x+4(a^2-3b^2)t))]\sec^2(2a(x+4(a^2-3b^2)t))}{2(a^2+b^2)} ,\\
&S_{12}=\frac{2ae^{-2b(x+4(3a^2-b^2)t)}(b-a\tan(2a(x+4(a^2-3b^2)t)))\sec(2a(x+4(a^2-3b^2)t))}{a^2+b^2},\\
&S_{21}=-\frac{2ae^{2b(x+4(3a^2-b^2)t)}(b+a\tan(2a(x+4(a^2-3b^2)t)))\sec(2a(x+4(a^2-3b^2)t))}{a^2+b^2}.
\end{align*}

Let us rewrite Eq. \eqref{eq4.5} in the vector form. Assume that $\pmb{p}=(p_1,p_2,p_3)$ with
\begin{eqnarray*}
p_1=\left(
\begin{array}{cc}
0 & -1\\
1 & 0\\
\end{array}\right), \quad p_2=\left(
\begin{array}{cc}
0 & 1\\
1 & 0\\
\end{array}\right),\quad p_3=\left(
\begin{array}{cc}
1 & 0\\
0 & -1\\
\end{array}\right)=\sigma_3.
\end{eqnarray*}
It is obvious that matrix $S$ is not Hermitian. It can be written as
\begin{eqnarray*}
S=\left(
\begin{array}{cc}
s_{3}& i(s_{2}-s_{1})\\
 i(s_{2}+s_{1}) &  -s_{3}\\
\end{array}\right),
\end{eqnarray*}
where components $s_j=s_j(x,t)$ of complex-valued vector $(s_1,s_2,s_3)$ are given by
\begin{equation}\begin{aligned}\label{eq4.23}
s_1&=\Delta \left(fg-f^{*}(-x,-t)g^{*}(-x,-t)\right),\quad
s_2=\Delta \left(fg+f^{*}(-x,-t)g^{*}(-x,-t)\right),\\
s_3&=i\Delta \left(gg^{*}(-x,-t)-ff^{*}(-x,-t)\right),\quad
\Delta=\left(ff^{*}(-x,-t)+gg^{*}(-x,-t)\right)^{-1}.
\end{aligned}\end{equation}
It is direct to check that $s_j$ satisfies
\begin{eqnarray*}
s_1(x,t)=-s_1^{*}(-x,-t),\quad s_2(x,t)=s_2^{*}(-x,-t),\quad s_3(x,t)=-s_3^{*}(-x,-t).
\end{eqnarray*}
By setting $S=U+iV$, where $U=i u_1p_1+i u_2 p_2+u_3p_3,V=i v_1p_1+i v_2 p_2+v_3p_3$,
then Eq.\eqref{eq4.5} is rewritten in the vector form
\begin{eqnarray}\begin{aligned}\label{eq4.22}
&\textbf{u}_t+\textbf{u}_{xxx}-\frac{3}{2}\left((\textbf{u}_x^2-\textbf{v}_x^2)\textbf{u}_x+
2\textbf{u}(\textbf{u}_x\textbf{u}_{xx}-\textbf{v}_x\textbf{v}_{xx})
-2\textbf{v}(\textbf{v}_x\textbf{u}_{xx}+\textbf{v}_{xx}\textbf{u}_x)\right)=0,\\
&\textbf{v}_t+\textbf{v}_{xxx}-\frac{3}{2}\left((\textbf{u}_x^2-\textbf{v}_x^2)\textbf{v}_x+
2\textbf{v}(\textbf{u}_x\textbf{u}_{xx}-\textbf{v}_x\textbf{v}_{xx})
+2\textbf{u}(\textbf{v}_x\textbf{u}_{xx}+\textbf{v}_{xx}\textbf{u}_x)\right)=0,
\end{aligned}
\end{eqnarray}
where real-valued vectors $\textbf{u}=(u_1,u_2,u_3),\textbf{v}=(v_1,v_2,v_3)$ satisfy $\textbf{u}^2-\textbf{v}^2=1, \textbf{u}\textbf{v}=0,$ and
$\textbf{a}\textbf{b}$ is defined by $\textbf{a}\textbf{b}=a_1b_1+a_2b_2-a_3b_3$. The relation between $u_j,v_j$ and
$s_j$ is $u_j=\textrm{Re}(s_j), v_j=\textrm{Im}(s_j)(j=1,2,3).$

We should remark here that though the matrix Eq.\eqref{eq4.5} has the same form as gauge equivalent version of the complex mKdV$^-$ equation, there exists a significant difference between the nonlocal complex mKdV$^-$ equation and the classical complex mKdV$^-$ equation. In fact, for the classical complex mKdV$^-$ equation, the form of matrix $G$ in \eqref{eq3}
\begin{eqnarray*}
G=\left(\begin{array}{cc}
f & g \\
f^* & -g^{*} \\
\end{array}\right)
\end{eqnarray*}
determines that $S=-iG^{-1}\sigma_3 G$ has the form
\begin{eqnarray*}
S=\frac{i}{\textrm{Re}(fg^*)}\left(
\begin{array}{cc}
-i \textrm{Im}(fg^*)& -|g|^2\\
-|f|^2 & i \textrm{Im}(fg^*)\\
\end{array}\right).
\end{eqnarray*}
Set $f=a(x,t)+ib(x,t), g=c(x,t)+id(x,t)$, then the matrix $S$ can be rewritten as
\begin{eqnarray*}
S=\left(
\begin{array}{cc}
s_{3}& i(s_{2}-s_{1})\\
 i(s_{2}+s_{1}) &  -s_{3}\\
\end{array}\right),
\end{eqnarray*}
where the vector $\textbf{S}$=$(s_{1},s_{2},s_{3})^T\in H^2$ in $R^{2+1}$, i.e., $s_1^2+s_2^2-s_3^2=-1,$ and $s_{j}(j=1,2,3)$ is given by
\begin{eqnarray*}
s_{1}=\frac{|g|^2-|f|^2}{2(ac+bd)},\qquad
s_{2}=\frac{-|f|^2-|g|^2}{2(ac+bd)},\qquad
s_{3}=\frac{bc-ad}{ac+bd}.
\end{eqnarray*}
Therefore Eq.\eqref{eq4.5} can be represented in vector form as
\begin{equation}\label{eq4.10}
\textbf{S}_t+\textbf{S}_{xxx}-\frac{3}{2}\left(\textbf{S}_{x}^2\textbf{S}_{x}+2\textbf{S}\textbf{S}_{x}\textbf{S}_{xx}\right)=0.
\end{equation}
We thus see that Eq.\eqref{eq4.22} is a general case of equation \eqref{eq4.10}.
%However, the matrix $S$ in \eqref{eq4.5} can be written as
%\begin{equation*}
%S=\left(
%\begin{array}{cc}
%s_1+i s_2 & -(s_3(-x,-t)-i s_4(-x,-t)) \\
%s_3+i s_4 & -(s_1+i s_2) \\
%\end{array}\right),
%\end{equation*}
%where
%\begin{eqnarray}\label{eq4.9}
%&&s_1s_1(-x,-t)+s_2s_2(-x,-t)+s_3s_3(-x,-t)+s_4s_4(-x,-t)=1,\nonumber\\
%&&s_1(-x,-t)s_2-s_2(-x,-t)s_1+s_3s_4(-x,-t)-s_4s_3(-x,-t)=0,
%\end{eqnarray}
%and $s_j (j=1,2,3,4)$ is given by
%\begin{equation*}
%s_1=\frac{s_{11}}{\Gamma},\quad s_2=\frac{s_{12}}{\Gamma},
%\quad s_3=\frac{s_{21}}{\Gamma},\quad s_4=\frac{s_{22}}{\Gamma},
%\end{equation*}
%with
%\begin{align}\nonumber
%\Gamma=&(a^2+b^2)(a^2(-x,-t)+b^2(-x,-t))+(c^2+d^2)(c^2(-x,-t)+d^2(-x,-t))\\\nonumber
%&+2(ac+bd)(a(-x,-t)c(-x,-t)+b(-x,-t)d(-x,-t))\\\nonumber
%&+2(bc-ad)(b(-x,-t)c(-x,-t)-a(-x,-t)d(-x,-t)),\\\nonumber
%s_{11}=&2(ac+bd)(a(-x,-t)d(-x,-t)-b(-x,-t)c(-x,-t))\\\nonumber
%&+2(bc-ad)(b(-x,-t)d(-x,-t)+a(-x,-t)c(-x,-t)),\\\nonumber
%s_{12}=&-(a^2+b^2)(a^2(-x,-t)+b^2(-x,-t))+(c^2+d^2)(c^2(-x,-t)+d^2(-x,-t)),\\\nonumber
%s_{21}=&-2(a^2+b^2)(da(-x,-t)+cb(-x,-t))-2(c^2+d^2)(bc(-x,-t)+ad(-x,-t)),\\\nonumber
%s_{22}=&2(a^2+b^2)(ca(-x,-t)-db(-x,-t))+2(c^2+d^2)(ac(-x,-t)-bd(-x,-t)).\nonumber
%\end{align}

\section{Solution and gauge equivalence of the nonlocal complex mKdV$^+$ equation }
In this section, we give the soliton solution and gauge equivalent structure for the nonlocal complex mKdV$^+$ equation:
\begin{equation}\label{eq4.1}
q_{t}+6 qq^*(-x,-t) q_x+q_{xxx}=0.
\end{equation}
%its Lax pair is
%\begin{equation}\label{eq4.2}
%\varphi_x=M \varphi,\quad \varphi_t=N \varphi,
%\end{equation}
%where
%\begin{subequations}
%\begin{align}
%&M=-i\lambda \sigma_3+Q,\\
%&N=-4i\lambda^3 \sigma_3+4\lambda^2Q-2i\lambda \sigma_3(Q^2-Q_x)+Q_x Q -Q Q_x-Q_{xx}+2Q^3,
%\end{align}
%\end{subequations}
%with
%\begin{align*}
%\sigma_3=\left(
%\begin{array}{cc}
%1 & 0 \\
%0 & -1 \\
%\end{array}\right),\quad
%Q=\left(
%\begin{array}{ccc}
%0 & q \\
%q^*(-x,-t) & 0 \\
%\end{array}\right).
%\end{align*}
%Suppose that $(f_1,f_2)^T=(f_1(x,t),f_2(x,t))^T$ is a solution of the eigenvalue equation \eqref{eq4.1} with $\lambda=\lambda_1$,
%then $(g_1,g_2)^T=(-f^{*}_2(-x,-t),f^{*}_1(-x,-t))^T$ is also the solution when $\lambda=-\lambda^{*}_1\triangleq \lambda_2$,
%\begin{equation}\label{}
%q^{[1]}=q-\frac{2i}{\Omega}(\lambda_1+\lambda^{*}_1)f_1f^{*}_2(-x,-t),
%\end{equation}
%where $\Omega=f_1f^{*}_1(-x,-t)+f_2f^{*}_2(-x,-t)$.
\textbf{Case 1.} For zero seed solution $q=0$, solving Eq. \eqref{eq2.1} gives the eigenfunctions
\begin{align}
f_1=e^{-i\lambda_1 x-4i\lambda_1^3t},\quad f_2=e^{i\lambda_1 x+4i\lambda_1^3t}.
\end{align}
So the new solution with $\lambda_1=a+ib(a\neq 0)$ is
\begin{align}\label{eq2.11}
\tilde{q}=-2ae^{2b(x+4(3a^2-b^2)t)}\csc(2a(x+4(a^2-3b^2)t)),
\end{align}
which has singularities at $\{(x,t)|x+4(a^2-3b^2)t=k\pi, k\in Z\}$.\\
\textbf{Case 2.} For nonzero seed solution
\begin{eqnarray}\label{eq2.12}
q=\rho e^{k(x-(k^2+6|\rho|^2)t)}\triangleq \rho e^{\Omega},
\end{eqnarray}
where $k$ is a real parameter and $\rho\neq0$ is a complex one. Solving Eq. \eqref{eq2.1} yields
\begin{align}\label{eq2.13}
&f_1=e^{\frac{k}{2}(x-(k^2+6|\rho|^2)t)}\left(c_1 e^{\frac{\xi}{2}(x+\eta t)}+c_2 e^{-\frac{\xi}{2}(x+\eta t)}\right),\\ \label{eq2.14}
&f_2=\frac{1}{2\rho}e^{-\frac{k}{2}(x-(k^2+6|\rho|^2)t)}\left(c_1(k+\xi+2i\lambda) e^{\frac{\xi}{2}(x+\eta t)}+c_2(k-\xi+2i\lambda) e^{-\frac{\xi}{2}(x+\eta t)}\right),
\end{align}
with $\xi=\sqrt{(k+2i\lambda_1)^2-4|\rho|^2}$ and $ \eta=4\lambda_1^2+2i \lambda_1 k-k^2-2|\rho|^2$, where $c_1$ and $c_2$ are nonzero complex parameters.

\begin{figure*}
\centering
\subfigure[] {\includegraphics[width=0.3\textwidth]{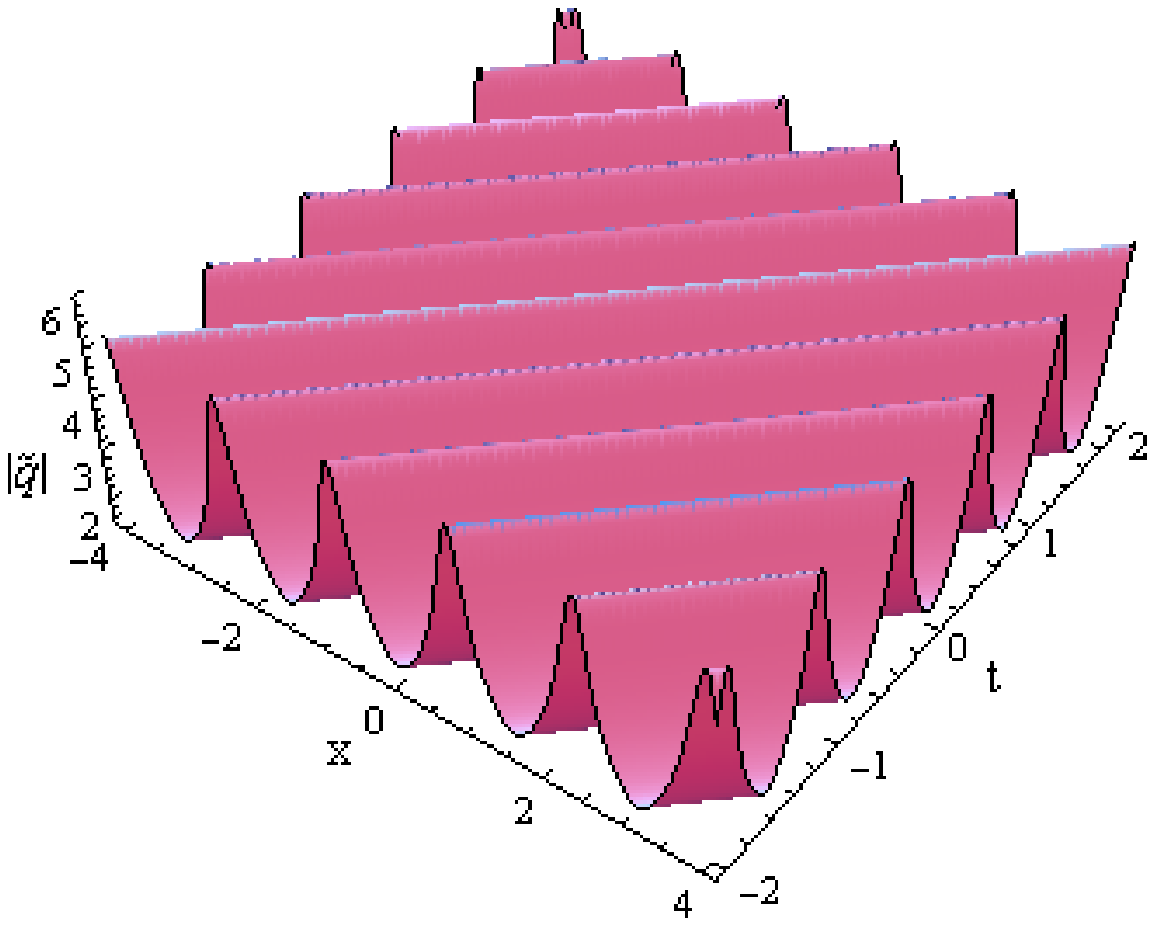}} \quad
\subfigure[$u=\tilde{q}\tilde{q}^{*}(-x,-t)$] {\includegraphics[width=0.3\textwidth]{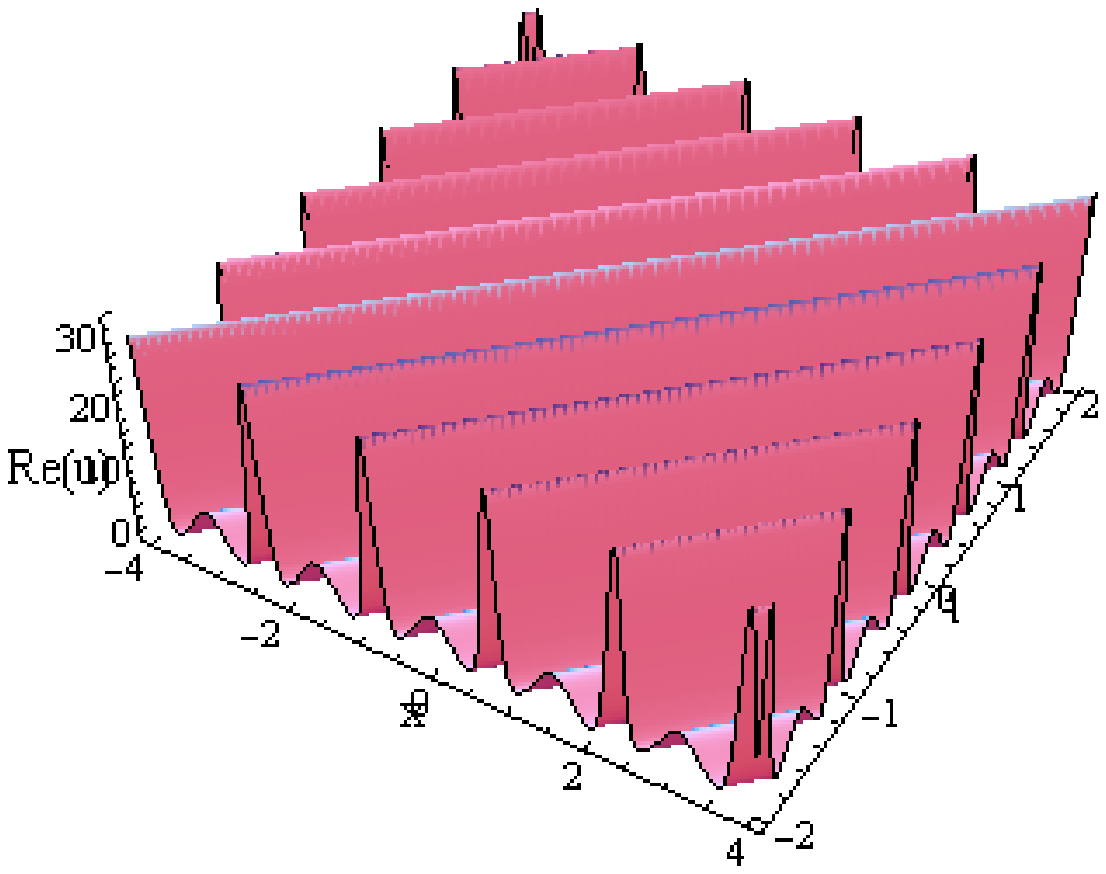}} \quad
\subfigure[$u=\tilde{q}\tilde{q}^{*}(-x,-t)$] {\includegraphics[width=0.3\textwidth]{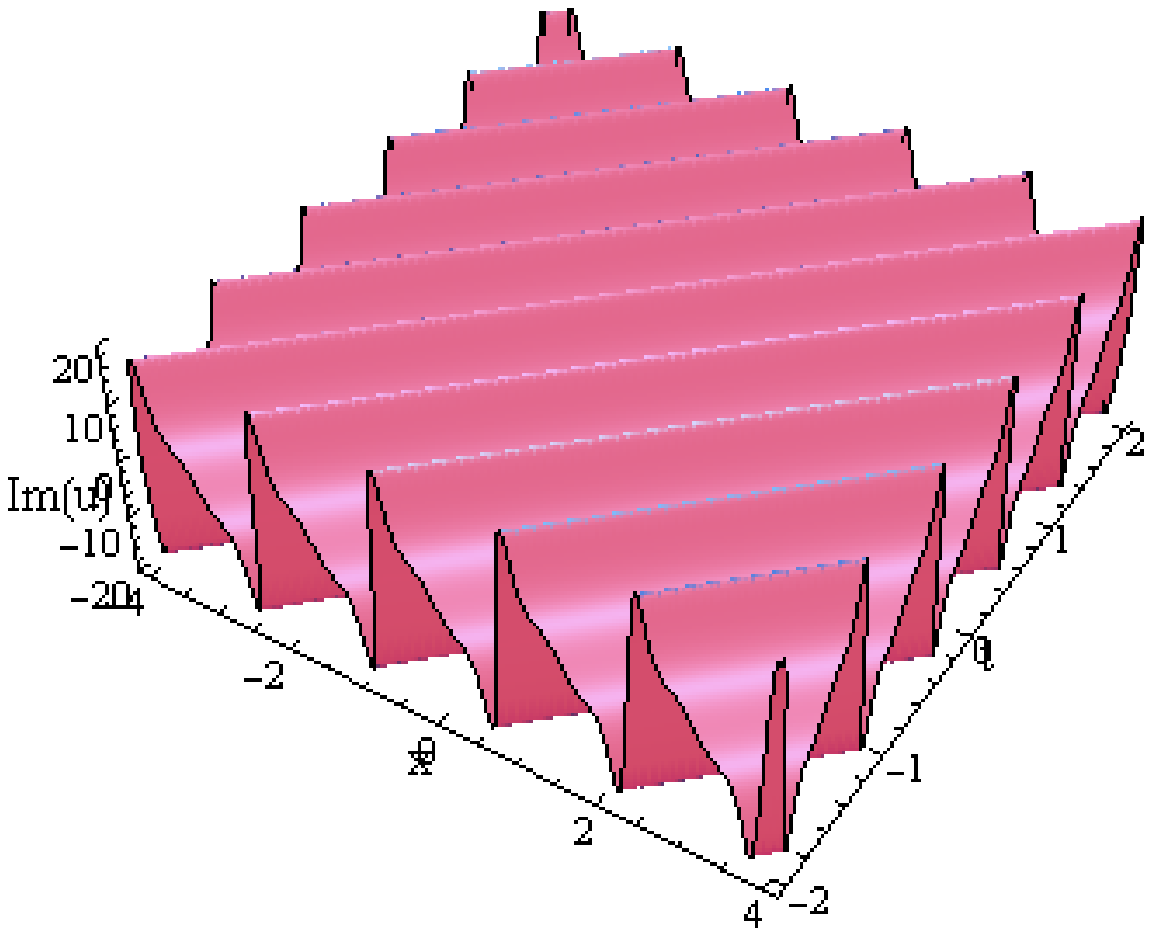}} \\
\subfigure[] {\includegraphics[width=0.28\textwidth]{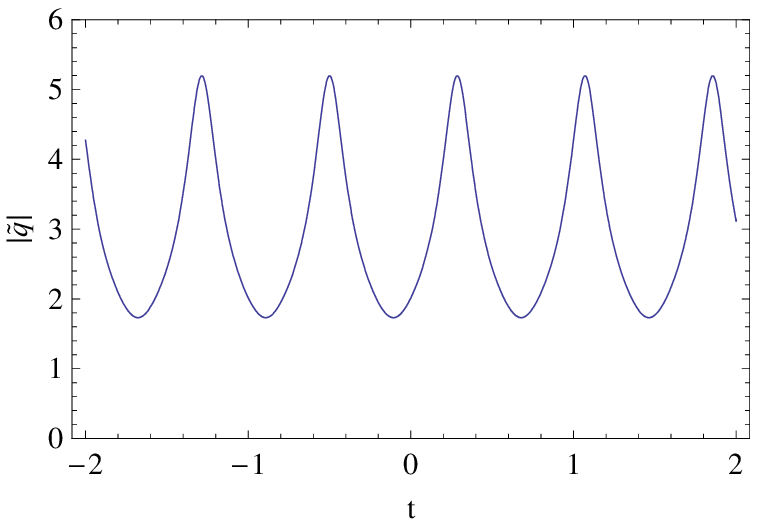}} \quad
\subfigure[] {\includegraphics[width=0.28\textwidth]{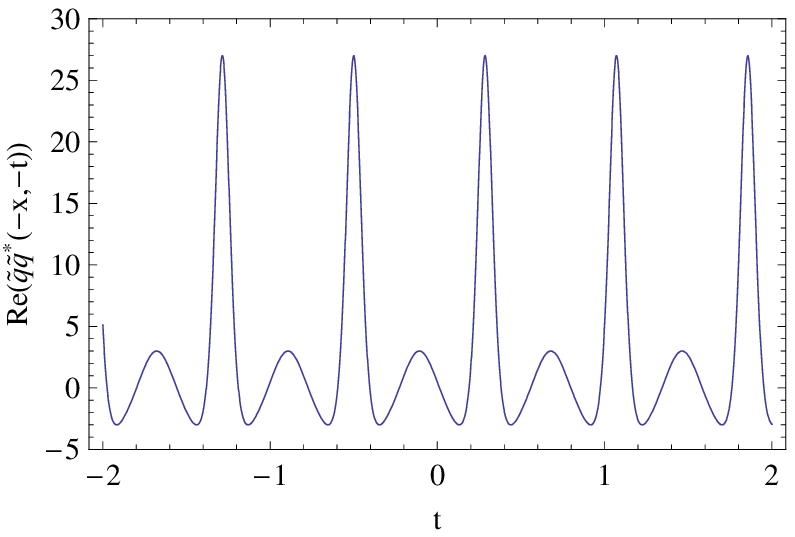}} \quad
\subfigure[] {\includegraphics[width=0.28\textwidth]{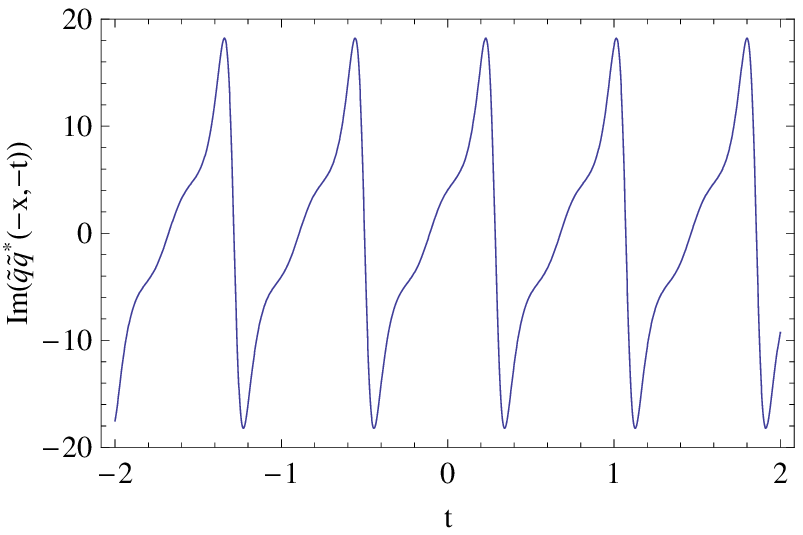}} \quad
\caption{\small{The space-time periodic solution with parameters $a=1,b=0,\gamma=1,\rho=\sqrt{3}i$. $T_{space}=\frac{\pi}{2}$ and $T_{time}=\frac{\pi}{4}$. (a)-(c) 3D plot: $|\tilde{q}|$, $\textrm{Re}(\tilde{q}\tilde{q}^*(-x,-t))$ and $\textrm{Im}(\tilde{q}\tilde{q}^*(-x,-t))$; (d-f) profiles at $x=-1$. }}
\label{Fig4}
\end{figure*}

\begin{figure*}
\centering
\subfigure[] {\includegraphics[width=0.3\textwidth]{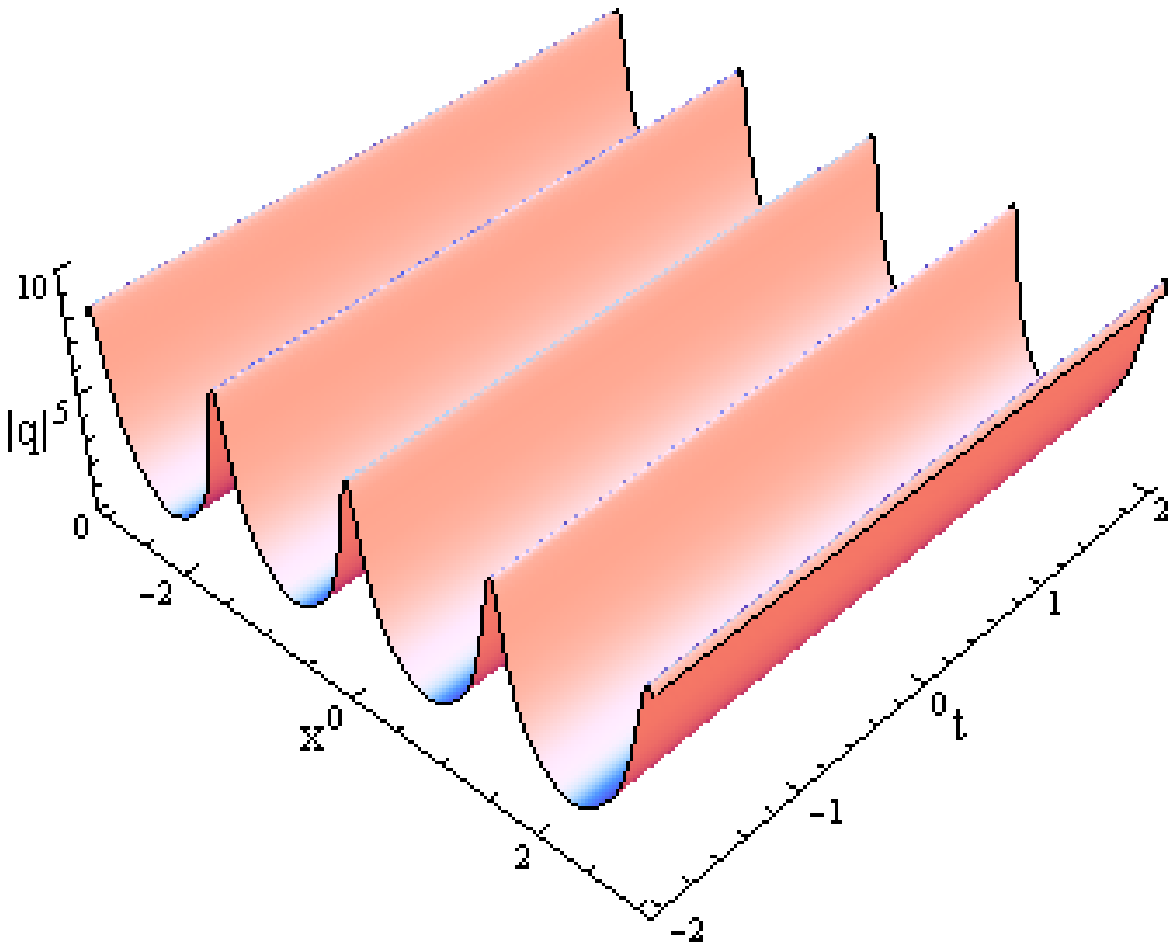}} \quad
\subfigure[] {\includegraphics[width=0.3\textwidth]{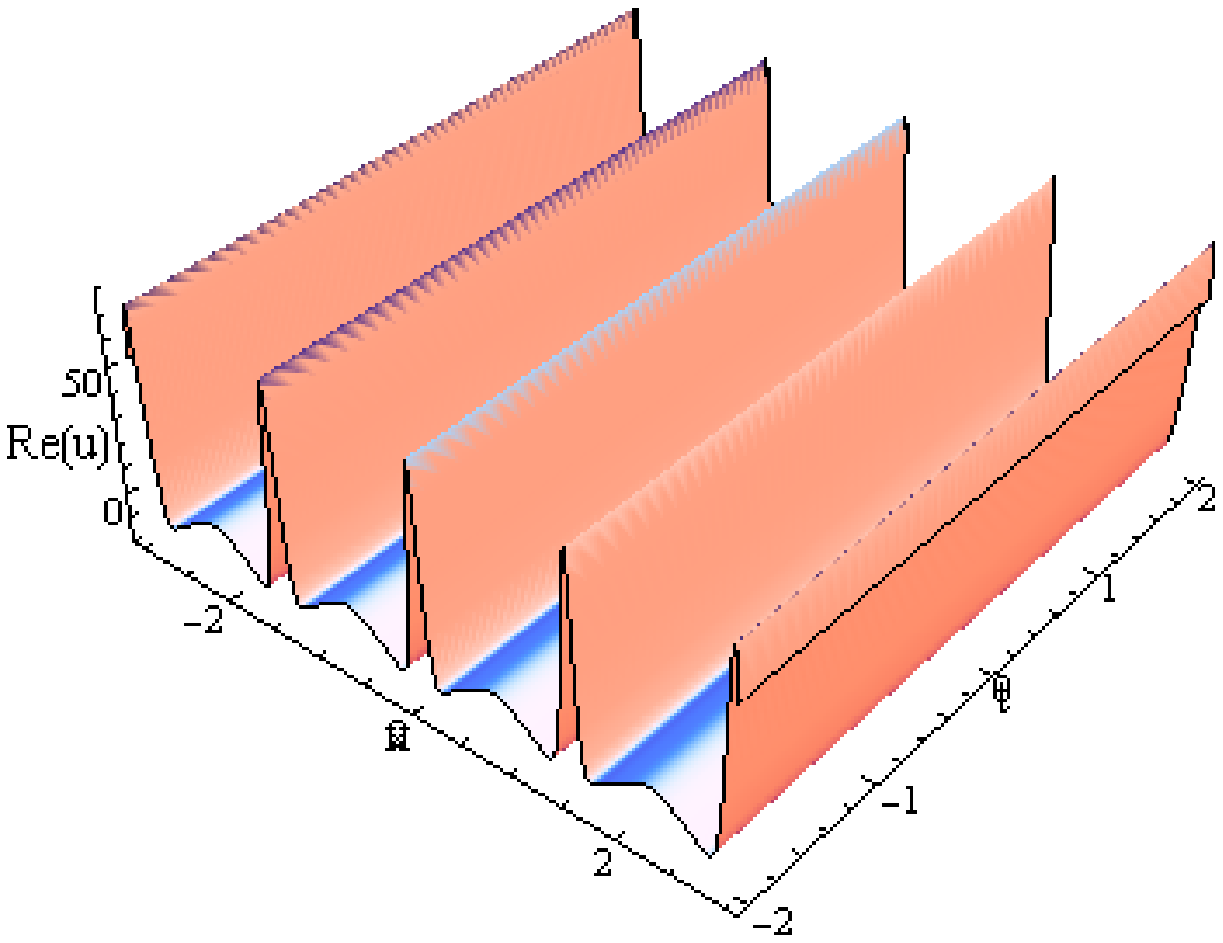}} \quad
\subfigure[] {\includegraphics[width=0.3\textwidth]{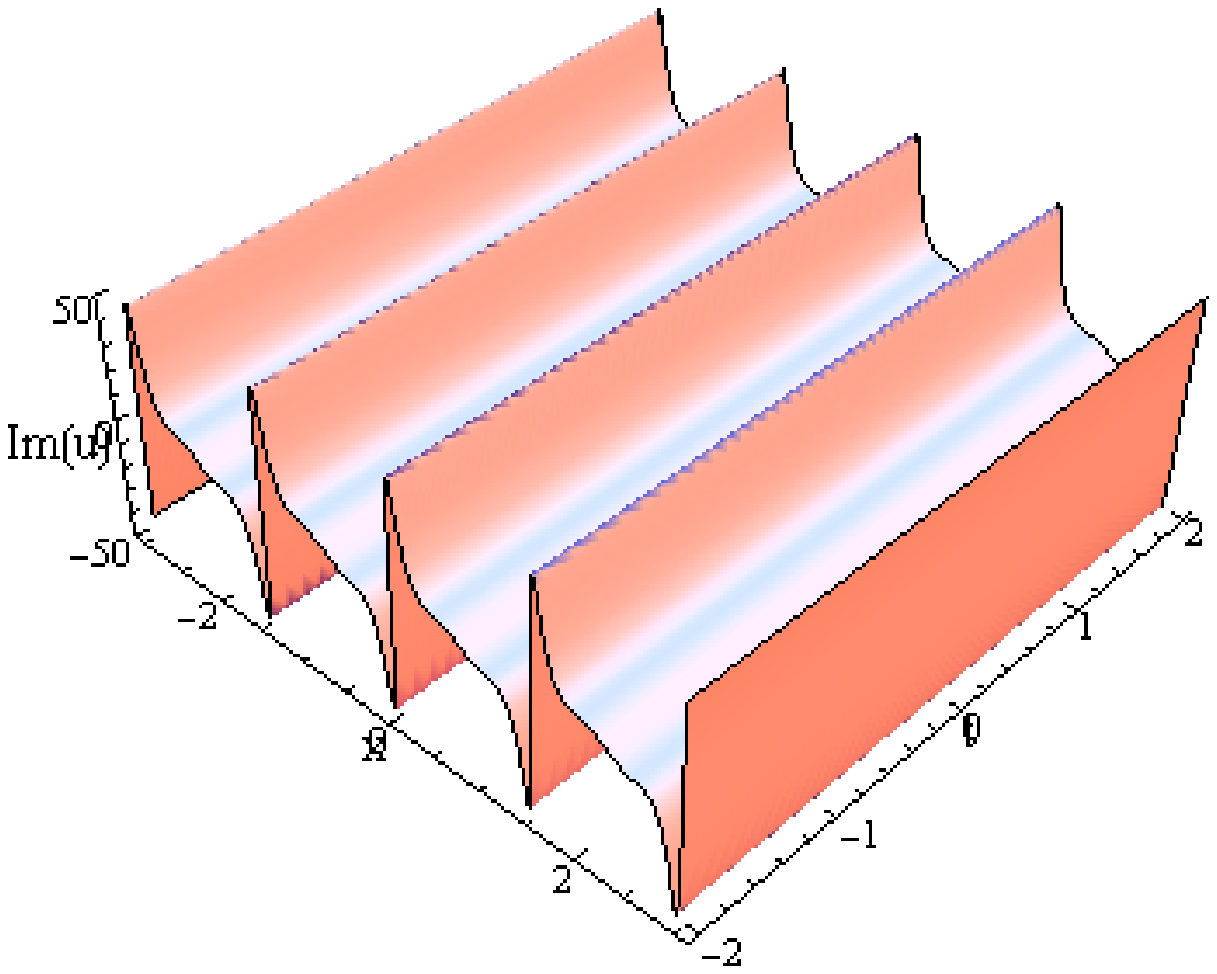}}
\caption{\small{Spatially periodic solution of the nonlocal complex mKdV$^+$ equation with $T_{space}=\frac{\sqrt{2}\pi}{3}$ as $a=\frac{\sqrt{6}}{2},b=0,\gamma=1,\rho=\sqrt{3}i$. (a)-(c) 3D plot: $|\tilde{q}|$, $u=\tilde{q}\tilde{q}^*(-x,-t)$. }}
\label{Fig5}
\end{figure*}

Let $k=2b$, then $\xi=2i\sqrt{a^2+|\rho|^2}\triangleq 2si, \eta=4a^2-12b^2-2|\rho|^2+12 i ab\triangleq \eta_1+i \eta_2$.
The eigenfunctions \eqref{eq2.13} and \eqref{eq2.14} can be rewritten as
\begin{align}\label{eq2.15}
&f_1=e^{b(x-2(2b^2+3|\rho|^2)t)}\left(c_1 e^{-s \eta_2 t+ is(x+\eta_1 t)}+c_2 e^{s \eta_2 t- is(x+\eta_1 t)}\right),\\ \label{eq2.16}
&f_2=e^{-b(x-2(2b^2+3|\rho|^2)t)}\left(\frac{ic_1(a+s)}{\rho} e^{-s \eta_2 t+ is(x+\eta_1 t)}+\frac{ic_2(a-s)}{\rho} e^{s \eta_2 t- is(x+\eta_1 t)}\right).
\end{align}
So we can get new solution
\begin{eqnarray}\label{eq2.17}
\tilde{q}=\rho e^{\Omega}\left(1-2a \frac{(a+s)e^{2is(x+\eta_1 t)}+\gamma^*(a-s)e^{-2s\eta_2t}+\gamma(a+s)e^{2s\eta_2t}+|\gamma|^2(a-s)e^{-2is(x+\eta_1 t)}}{a(a+s) e^{2is(x+\eta_1 t)}-\gamma^*|\rho|^2e^{-2s\eta_2t}-\gamma|\rho|^2e^{2s\eta_2t}+a|\gamma|^2(a-s) e^{-2is(x+\eta_1 t)}}\right).
\end{eqnarray}
As $k>0$ and $\eta_2>0$,
\begin{align}
&\tilde{q}=\rho e^{\Omega}\left(1+\frac{2a(a+s)}{|\rho|^2}\right)\sim 0, \qquad t\rightarrow +\infty,\\
&\tilde{q}=\rho e^{\Omega}\left(1+\frac{2a(a-s)}{|\rho|^2}\right)\sim +\infty , \quad t\rightarrow -\infty.
\end{align}
Note that $(a+s)(a-s)=-|\rho|^2$, as $b=0$ and $a(a+s)+a(a-s)|\gamma|^2\neq \pm|\rho|^2(\gamma+\gamma^*)$, the solution \eqref{eq2.17} is always smooth and has the periods $T_{space}=\frac{\pi}{s}$ and $T_{time}=\frac{\pi}{s\eta_1}$ in space and time, respectively. We choose the parameters $a=1,b=0,\gamma=1,\rho=\sqrt{3}i$, the periodic solutions are shown in Fig.\ref{Fig4}. When $b=0$ and $2a^2=|\rho|^2$, a spatially periodic solution with period $T_{space}=\frac{\pi}{s}$ of the nonlocal complex mKdV$^+$ equation is described in Fig.\ref{Fig5}.

We remark here that these solutions are different from the solution
\begin{eqnarray}\begin{aligned}
q=-\frac{(\eta+\bar{\eta})e^{i\bar{\theta}}e^{-2\bar{\eta}x+8\bar{\eta}^3t}}{1+e^{i(\theta+\bar{\theta})}e^{-2\eta x+8\eta^3 t-2\bar{\eta}x+8\bar{\eta}^3t}},
\end{aligned}
\end{eqnarray}
which was given in Ref.\cite{0.1}.

Next, let us seek to the gauge equivalent structure of the nonlocal complex mKdV$^+$ equation. For the nonlocal complex mKdV$^+$ equation \eqref{eq4.1}, using \eqref{eq2.1} and gauge transformation \eqref{gt}, we can get
\begin{subequations}
\begin{align}\label{eq4}
&\tilde{M}=-i\lambda G^{-1}\sigma_3 G \triangleq -i\lambda S, \\
&\tilde{N}=-4i\lambda^3S+2\lambda^2 SS_x+i\lambda(S_{xx}+\frac{3}{2}SS_x^2),
\end{align}
\end{subequations}
in which we have used the following important identities
\begin{equation}
SS_x=2G^{-1}QG, \quad SS_x^2=-4G^{-1}\sigma_3Q^2G,\quad S_{xx}+SS_x^2=2G^{-1}\sigma_3Q_xG.
\end{equation}
The integrability condition $\tilde{M}_t-\tilde{N}_x+[\tilde{M},\tilde{N}]=0$ for new linear equations
\begin{equation}\label{eq5}
\tilde{\varphi}_x=\tilde{M} \tilde{\varphi},\quad \tilde{\varphi}_t=\tilde{N} \tilde{\varphi},
\end{equation}
leads to
\begin{equation}\label{eq6}
S_t+S_{xxx}+\frac{3}{2}\left(S_x^3+SS_{xx}S_{x}+SS_{x}S_{xx}\right)=0,
\end{equation}
where $S^2=I, \textrm{Tr}S=0$.

The structure of the matrices $M(0)$ and $N(0)$ implies that the solution $G$ of Eq. \eqref{eq3} has the form
\begin{equation}\label{eq7}
G=\left(
\begin{array}{cc}
f & g^{*}(-x,-t) \\
g & f^{*}(-x,-t) \\
\end{array}\right).
\end{equation}
Hence the structure of the matrix $S$ can be given by
\begin{equation}\label{eq8}
S=G^{-1}\sigma_3 G=\frac{1}{\Delta}\left(
\begin{array}{cc}
ff^{*}(-x,-t)+gg^{*}(-x,-t) & 2f^{*}(-x,-t)g^{*}(-x,-t) \\
-2fg & -ff^{*}(-x,-t)-gg^{*}(-x,-t) \\
\end{array}\right),
\end{equation}
where $\Delta=ff^{*}(-x,-t)-gg^{*}(-x,-t)$.\\
For the given solution \eqref{eq2.11} of the nonlocal complex mKdV$^+$ equation, solving the Eq. \eqref{eq3} yields
\begin{eqnarray}\label{eq9}
G=\left(
\begin{array}{cc}
-\frac{ia e^{2b(x+4(3a^2-b^2)t)}\csc(2a(x+4(a^2-3b^2)t))}{\sqrt{a^2+b^2}} & i \frac{-b- a\cot(2a(x+4(a^2-3b^2)t)}{\sqrt{a^2+b^2}} \\\\
-i \frac{-b+a\cot(2a(x+4(a^2-3b^2)t)}{\sqrt{a^2+b^2}} & -\frac{ia e^{-2b(x+4(3a^2-b^2)t)}\csc(2a(x+4(a^2-3b^2)t))}{\sqrt{a^2+b^2}} \\
\end{array}\right).
\end{eqnarray}
Thus we can derive the exact solution of Eq. \eqref{eq6}
\begin{equation*}
S=\left(
\begin{array}{cc}
S_{11} & S_{12} \\
S_{21} & -S_{11} \\
\end{array}\right),
\end{equation*}
where
\begin{eqnarray*}
\begin{aligned}
&S_{11}=-\frac{[3a^2-b^2+(a^2+b^2)\cos(4a(x+4(a^2-3b^2)t))]\csc^2(2a(x+4(a^2-3b^2)t))}{2(a^2+b^2)} ,\\
&S_{12}=-\frac{2ae^{-2b(x+4(3a^2-b^2)t)}(b+a\cot(2a(x+4(a^2-3b^2)t)))\csc(2a(x+4(a^2-3b^2)t))}{a^2+b^2},\\
&S_{21}=\frac{2ae^{2b(x+4(3a^2-b^2)t)}(-b+a\cot(2a(x+4(a^2-3b^2)t)))\csc(2a(x+4(a^2-3b^2)t))}{a^2+b^2}.
\end{aligned}
\end{eqnarray*}

Next, we rewrite Eq.\eqref{eq6} in the vector form. As mentioned in \cite{11}, $S$ can be expressed explicitly as
\begin{eqnarray*}
S=\left(
\begin{array}{cc}
s_{3}& s_{1}-is_{2}\\
 s_{1}+is_{2} &  -s_{3}\\
\end{array}\right),
\end{eqnarray*}
where complex-valued functions $s_j(j=1,2,3)$ are determined by \eqref{eq8} and satisfy
\begin{eqnarray*}
s_1(x,t)=-s_1^{*}(-x,-t),\quad s_2(x,t)=-s_2^{*}(-x,-t),\quad s_3(x,t)=s_3^{*}(-x,-t).
\end{eqnarray*}
One can check $S=\sigma_3S^\dag(-x,-t)\sigma_3$. Set $S=U+iV$, where Hermitian matrices $U$ and $V$ are
\begin{eqnarray*}U=\frac{S+S^\dag}{2},\quad V=i\frac{S^\dag-S}{2},\end{eqnarray*}
and have Pauli matrix representation $U=\textbf{u}\cdot\pmb{\sigma},V=\textbf{v}\cdot\pmb{\sigma}$, where
$\pmb{\sigma}=(\sigma_1,\sigma_2,\sigma_3)$ with $\sigma_j$
\begin{eqnarray*}
\sigma_1=\left(
\begin{array}{cc}
0 & 1\\
1 & 0\\
\end{array}\right), \quad \sigma_2=\left(
\begin{array}{cc}
0 & -i\\
i & 0\\
\end{array}\right),\quad \sigma_3=\left(
\begin{array}{cc}
1 & 0\\
0 & -1\\
\end{array}\right).
\end{eqnarray*}
Then the vector form of Eq.\eqref{eq6} is given by
\begin{eqnarray}\begin{aligned}\label{eq10}
&\textbf{u}_t+\textbf{u}_{xxx}+\frac{3}{2}\left((\textbf{u}_x^2-\textbf{v}_x^2)\textbf{u}_x+
2\textbf{u}(\textbf{u}_x\textbf{u}_{xx}-\textbf{v}_x\textbf{v}_{xx})
-2\textbf{v}(\textbf{v}_x\textbf{u}_{xx}+\textbf{v}_{xx}\textbf{u}_x)\right)=0,\\
&\textbf{v}_t+\textbf{v}_{xxx}+\frac{3}{2}\left((\textbf{u}_x^2-\textbf{v}_x^2)\textbf{v}_x+
2\textbf{v}(\textbf{u}_x\textbf{u}_{xx}-\textbf{v}_x\textbf{v}_{xx})
+2\textbf{u}(\textbf{v}_x\textbf{u}_{xx}+\textbf{v}_{xx}\textbf{u}_x)\right)=0,
\end{aligned}
\end{eqnarray}
where real-valued vectors $\textbf{u}=(u_1,u_2,u_3)$ and $\textbf{v}=(v_1,v_2,v_3)$ satisfy $\textbf{u}\textbf{u}-\textbf{v}\textbf{v}=1, \textbf{u}\textbf{v}=0$, and $\textbf{a}\textbf{b}=a_1b_1+a_2b_2+a_3b_3$.

We remark here that the matrix $S$ in \eqref{eq6} differs from that for the classical complex mKdV$^+$ equation. In fact, for the complex mKdV$^+$ equation, the solution $G$ of Eq.\eqref{eq3} has the form
\begin{eqnarray*}
G=\left(\begin{array}{cc}
f & -g^{*} \\
g & f^{*}\\
\end{array}\right),
\end{eqnarray*}
and thus the matrix $S$ has the form,
\begin{eqnarray*}
S=\frac{1}{|f|^2+|g|^2}\left(
\begin{array}{cc}
|f|^2-|g|^2& -2f^{*}g^{*}\\
-2fg & |g|^2-|f|^2 \\
\end{array}\right).
\end{eqnarray*}
Assume that $f=a+ib, g=c+id$, then the matrix $S$ can be written as
\begin{eqnarray*}
S=\left(
\begin{array}{cc}
s_{3}& s_{1}-is_{2}\\
 s_{1}+is_{2} &  -s_{3}\\
\end{array}\right),
\end{eqnarray*}
where the real-valued vector $\textbf{S}$=$\{(s_{1},s_{2},s_{3})|s_{1}^2+s_{2}^2+s_{3}^2=1\}$ in $R^3$, and $s_{j}, j=1,2,3$ are given by
\begin{eqnarray*}
s_{1}=\frac{2(bd-ac)}{|f|^2+|g|^2},\qquad
s_{2}=\frac{-2(bc+ad)}{|f|^2+|g|^2},\qquad
s_{3}=\frac{|f|^2-|g|^2}{|f|^2+|g|^2}.
\end{eqnarray*}
Eq.\eqref{eq6} can be rewritten in vector form as
\begin{equation}\label{eq66}
\textbf{S}_t+\textbf{S}_{xxx}+\frac{3}{2}\left(\textbf{S}_{x}^2\textbf{S}_{x}+2\textbf{S}\textbf{S}_{x}\textbf{S}_{xx}\right)=0.
\end{equation}
In brief, we can see that there exist great differences in the properties between the nonlocal complex mKdV and the classical complex mKdV equation.
\section{Conclusions}
In this paper, we have shown that, under the gauge transformations, the nonlocal complex mKdV equation is gauge equivalent to a spin equation. From the gauge equivalence, we can see that the properties between the nonlocal complex mKdV and the classical complex mKdV equation have great differences. By constructing the Darboux transformation, we have obtained distinct kinds of exact solutions including dark soliton, W-type and M-type soliton and periodic solutions of the nonlocal complex mKdV equation. We hope to investigate the discrete integrable version of the nonlocal complex mKdV equation in the future.

\vskip 15pt \noindent {\bf
Acknowledgements} \vskip 12pt

The work of ZNZ is supported by the National Natural Science
Foundation of China (NNSFC)under grants 11271254, 11428102 and 11671255, and
in part by the Ministry of Economy and Competitiveness of Spain under
contracts MTM2012-37070 and MTM2016-80276-P, and the that of SFS is supported by NNSFC under grant 11371323.

\small{

}
\end{document}